\documentclass[%
 reprint,
superscriptaddress,
 amsmath,amssymb,
 aps,
 prl
]{revtex4-2}

\usepackage{graphicx}
\usepackage{dcolumn}
\usepackage{bm}
\usepackage{lipsum}  
\usepackage{amsmath}
\usepackage{mathrsfs}  
\usepackage{comment}
\usepackage{mathtools}
\usepackage{bigints}
\usepackage{algorithm2e}

\usepackage{caption}
\SetArgSty{text}
\RestyleAlgo{ruled}
\def\mathclap#1{\text{\hbox to 0pt{\hss$\mathsurround=0pt#1$\hss}}}

\begin{document}

\title{Learning Neural Free-Energy Functionals with Pair-Correlation Matching}

\date{\today}

\author{Jacobus Dijkman}
\affiliation{Van 't Hoff Institute for Molecular Sciences, University of Amsterdam, The Netherlands}
\affiliation{Informatics Institute, University of Amsterdam, The Netherlands}   
\author{Marjolein Dijkstra}
\affiliation{Soft Condensed Matter \& Biophysics, Debye Institute for Nanomaterials Science, Utrecht University, The Netherlands}
\author{Ren\'{e} van Roij}
\affiliation{Institute for Theoretical Physics, Utrecht University, The Netherlands}
\author{\\Max Welling}
\affiliation{Informatics Institute, University of Amsterdam, The Netherlands}
\author{Jan-Willem van de Meent}
\affiliation{Informatics Institute, University of Amsterdam, The Netherlands}
\author{Bernd Ensing}
\affiliation{Van 't Hoff Institute for Molecular Sciences, University of Amsterdam, The Netherlands}
\affiliation{AI4Science Laboratory, University of Amsterdam, The Netherlands}


\begin{abstract}
The intrinsic Helmholtz free-energy functional, the centerpiece of classical density functional theory (cDFT), is at best only known approximately for 3D systems. Here we introduce a method for learning a neural-network approximation of this functional by exclusively training on a dataset of radial distribution functions, circumventing the need to sample costly heterogeneous density profiles in a wide variety of external potentials. For a supercritical 3D Lennard-Jones system, we demonstrate that the learned neural free-energy functional accurately predicts planar inhomogeneous density profiles under various complex external potentials obtained from simulations.
\end{abstract}


\maketitle


\begin{figure*}[t]
\includegraphics[width=0.89\textwidth]{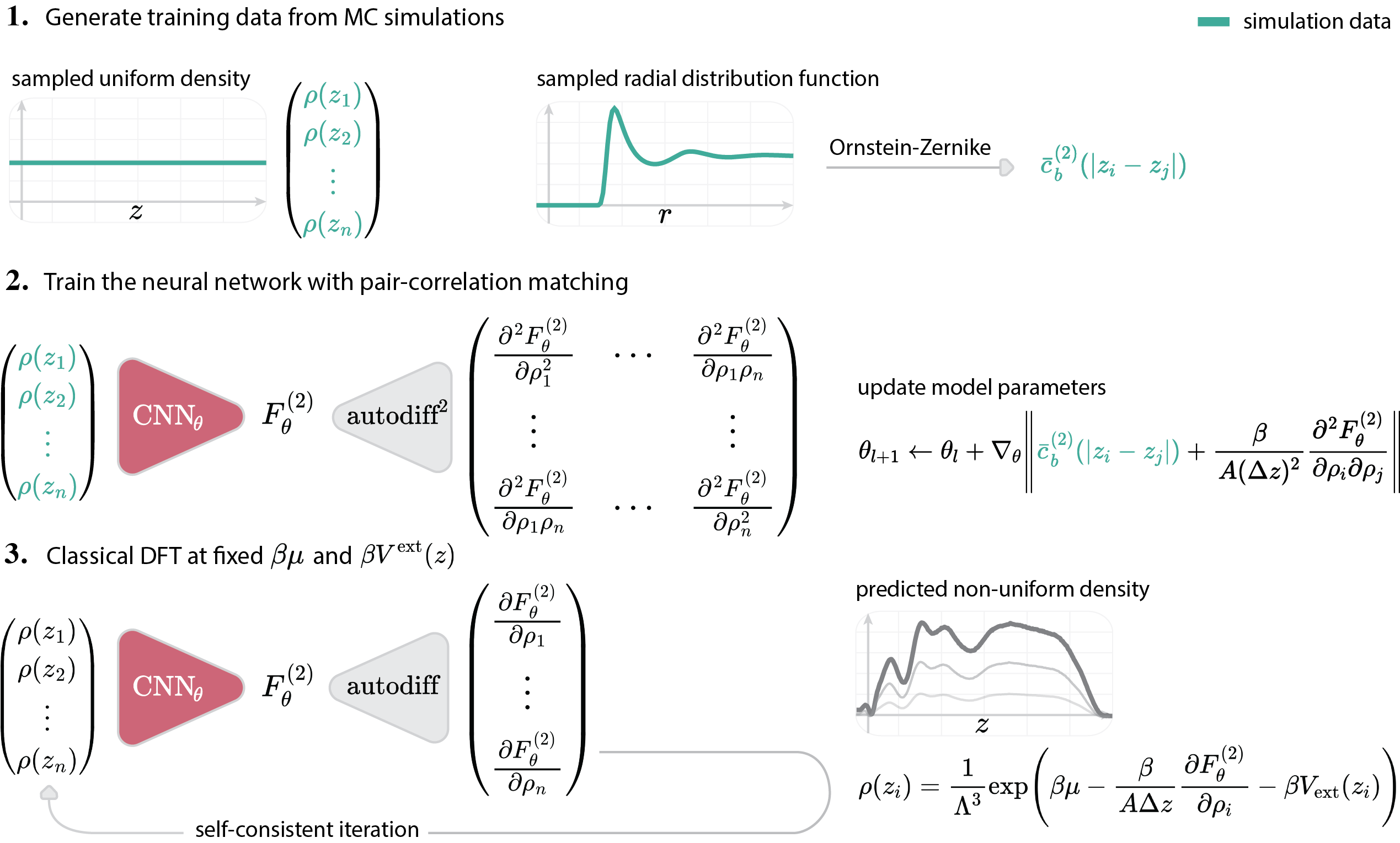}
\caption{\small \textbf{1.} Bulk densities in planar geometry $\rho(z_i)=\rho_b$ and radial distribution functions $g(r)$ are sampled from Monte Carlo simulations of homogeneous bulk systems of Lennard-Jones particles. Each $g(r)$ is converted to the second functional derivative of the excess free energy $\delta^2 \mathscr{F}_{\text{exc}} / \delta \rho(z_i) \delta \rho(z_j)$ by employing the Ornstein-Zernike equation. \textbf{2.} Through automatic differentiation (autodiff$^2$), the neural functional $F_{\theta}^{(2)}$ is optimized to fit the Hessian of the model output with respect to input density profiles to $\delta^2 \mathscr{F}_{\text{exc}} / \delta \rho(z_i) \delta \rho(z_j)$. \textbf{3.} The optimized model can then be applied in cDFT to obtain non-uniform equilibrium density profiles through automatic differentiation (autodiff) and the free energy $F_{\theta}^{(2)}$ for a system of Lennard-Jones particles subjected to arbitrary external potentials.}
\label{fig:overview}
\end{figure*}

Inhomogeneous many-body systems play a profound role in both science and technology, with examples spanning from p-n junctions in  semiconducting solid-state devices to phenomena like hydrogen bubbles in electrolysers, or gas adsorption in porous materials \cite{soaresClassicalDensityFunctional2023, guoClassicalDensityFunctional2016, guoFastScreeningPorous2018, guoScreeningPorousMaterials2020,fuClassicalDensityFunctional2015, liuClassicalDensityFunctional2016}. Density functional theory (DFT) is a powerful theoretical framework to describe the thermodynamic equilibrium properties and the structure of such systems, relying solely on the one-body density profile  $\rho({\bf r})$ \cite{evansNatureLiquidvapourInterface1979}. Classical DFT (cDFT) relies on the existence of an excess free-energy functional $\mathscr{F}_{\text{exc}}[\rho]$, which describes the non-ideal contribution to the total intrinsic free-energy functional $\mathscr{F}[\rho]$ and encompasses the inter-particle interactions. However, the main bottleneck of cDFT is that this functional is in general unknown, and  hence, one has to rely on developing accurate approximations for the excess free-energy functionals. 

Historically, the field of cDFT has emphasized development of analytical approximations for $\mathscr{F}_{\text{exc}}[\rho]$, often based on thermodynamics and (direct) pair correlations derived from \textit{approximate} closures of the Ornstein-Zernike equation of bulk systems \cite{hansenTheorySimpleLiquids2013}. For instance, the highly successful Fundamental Measure Theory (FMT) for hard spheres is deeply connected to the Percus-Yevick closure \cite{rothFundamentalMeasureTheory2010}, and many functionals for systems with soft Van der Waals or Coulombic interactions build on mean-field and mean-spherical approximations \cite{catsPrimitiveModelElectrolytes2021, hansenTheorySimpleLiquids2013}.

In recent years, there has been a resurgence of cDFT developments facilitated by machine learning (ML) methods, which employ virtually \textit{exact} thermodynamic and structural data obtained from explicit many-body simulations to learn data-driven representations of the excess free-energy functional $\mathscr{F}_{\text{exc}}[\rho]$. In the classical regime, the first machine-learned cDFTs focused on  supercritical Lennard-Jones fluids, for which explicit approximate functional forms for $\mathscr{F}_{\text{exc}}[\rho]$ were fitted to density profiles in external fields obtained from simulations, both for 1D  \cite{linClassicalDensityFunctional2019} and  3D systems in planar geometry \cite{catsMachinelearningFreeenergyFunctionals2021}. Recent work, once again leveraging simulations of density profiles in a variety of external potentials, has shown that a neural approximation of the functional derivative $\delta \mathscr{F}_{\text{exc}}/\delta \rho$ for hard-sphere systems outperforms FMT \cite{sammullerNeuralFunctionalTheory2023a} in accurately  estimating inhomogeneous density profiles.

In this Letter, we introduce a neural free-energy functional that we train using \textit{pair-correlation matching}, a novel optimization objective that matches the Hessian of the neural approximation to pair correlations of particles. We show that pair-correlation matching yields a neural functional that accurately predicts the excess free energy for a one-component system of interacting particles. The differentiable nature of this neural functional allows access to, and learning from, various structural and thermodynamic properties by utilizing the first and second functional derivatives of $\mathscr{F}_{\text{exc}}[\rho]$. Unlike previous ML approaches to cDFT \cite{linClassicalDensityFunctional2019,catsMachinelearningFreeenergyFunctionals2021,sammullerNeuralFunctionalTheory2023a,simonMachineLearningDensity2024}, our neural functional is trained by directly learning particle correlations from radial distribution functions sampled from short simulations of homogeneous bulk systems (illustrated in Fig.~\ref{fig:overview}), rather than inferring them from a costly dataset of inhomogeneous densities. We demonstrate that this neural free-energy functional can be applied in the cDFT framework to achieve accurate estimates for inhomogeneous density profiles in external fields without ever having seen any inhomogeneous densities during training. Simultaneously, the neural free-energy functional provides accurate estimates of the excess free energy.

Central to classical DFT is the grand canonical equilibrium density 
\begin{equation}
    \rho_0(\mathbf{r})= \frac{1}{\Lambda^{3}}\exp \left(\beta\mu-\left.\beta \frac{\delta \mathscr{F}_{\text {exc}}[\rho]}{\delta \rho(\mathbf{r})}\right|_{\rho=\rho_0}\hspace{-5mm}-\beta V_{\text {ext }}(\mathbf{r})\right),
\label{eq:equil_dens}
\end{equation}
with $\beta=1/k_BT$, $\Lambda$ the thermal wavelength, $\mu$ the chemical potential and $V_{\text{ext}}(\mathbf{r})$ the external potential. This self-consistency relation can be leveraged to find $\rho_0(\mathbf{r})$ through recursive Picard iteration \cite{rothIntroductionDensityFunctional2006, edelmannNumericalEfficientWay2016, mairhoferNumericalAspectsClassical2017}. Since $\rho_0(\mathbf{r})$ is dependent on $\mathscr{F}_{\text{exc}}[\rho]$ through the first functional derivative $\delta \mathscr{F}_{\text {exc}}[\rho]/\delta \rho(\mathbf{r})$, previous approaches to leveraging machine learning for cDFT \cite{linClassicalDensityFunctional2019, catsMachinelearningFreeenergyFunctionals2021, sammullerNeuralFunctionalTheory2023a} involve training a model to capture $\delta \mathscr{F}_{\text{exc}}/\delta \rho (\mathbf{r})$, which can be derived from sampled inhomogeneous equilibrium density profiles and employing Eq.(\ref{eq:equil_dens}).


\begin{figure*}
    \begin{center}
    \includegraphics[width=1\textwidth]{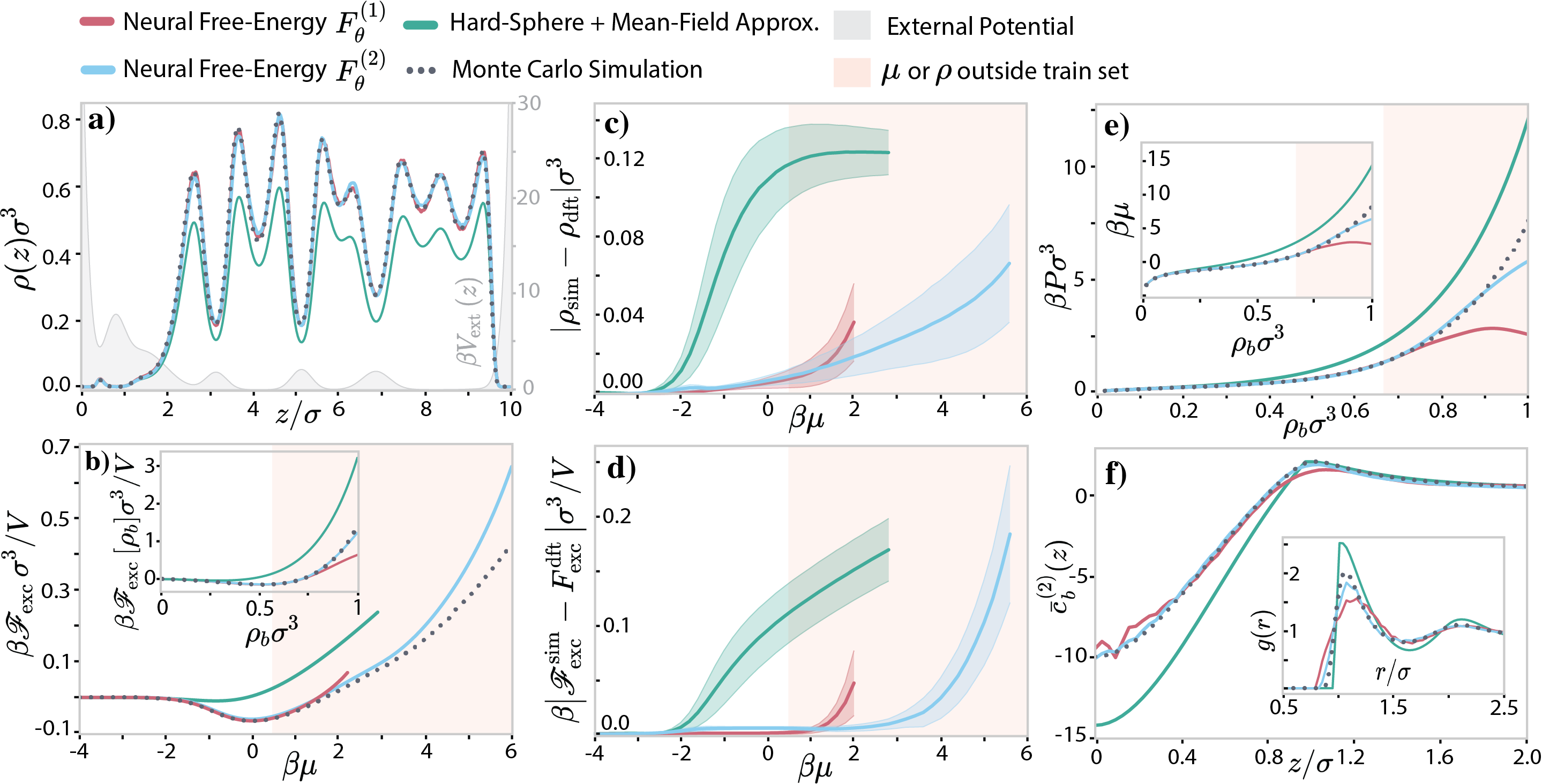}
    \end{center}
    \vspace{-1.5em} 
    \caption{\small Evaluation of neural free-energy functionals $F_\theta^{(1)}$ and $F_\theta^{(2)}$, where $F_\theta^{(1)}$ is optimized by matching inhomogeneous one-body densities and $F_\theta^{(2)}$ by pair-correlation matching in the homogeneous bulk. \textbf{a)} Density profiles of a Lennard-Jones system in a planar geometry characterized by an external potential (shown in gray) at a chemical potential of $\beta\mu = 0$ obtained from DFT using  $F_\theta^{(1)}$ and $F_\theta^{(2)}$ and the mean-field approximation $F_{\text{exc}}^{\text{MF}}$, along with the simulated density profile. \textbf{b)} Comparison of the free-energy estimates using $F_\theta^{(1)}$, $F_\theta^{(2)}$, and $F_{\text{exc}}^{\text{MF}}$ for the specific external potential shown in a), evaluated across the range $-4<\beta\mu <6$. Inset shows the bulk excess free-energy obtained from DFT and simulations across a range of bulk densities $0<\rho_b\sigma^3 < 1$. \textbf{c)} Mean absolute error $\frac{1}{n} \sum\left|\rho_{\text {sim}}\left(z_i\right)-\rho_{\text{dft}}\left(z_i\right)\right|$ between density profiles obtained from DFT using $F_\theta^{(1)}$,  $F_\theta^{(2)}$, and $F_{\text{exc}}^{\text{MF}}$ and densities sampled from simulation. \textbf{d)} Absolute error of the excess free-energy from DFT and simulations. \textbf{c)/d)} Data is shown for 150 distinct external potentials, evaluated across the range $-4<\beta\mu <6$, with steps of $\Delta \beta\mu = 0.2$. The area of the mean $\pm$ standard deviation is colored. The error is shown up to the point where the Picard iterations stop to converge to a solution within 1000 iterations. \textbf{e)} The bulk pressure and chemical potential obtained from DFT and simulations across a range of bulk densities $0<\rho_b\sigma^3 < 1$. \textbf{f)} The laterally integrated direct correlation function $\bar{c}_b^{(2)}(z)$ at $\rho_b\sigma^3 = 0.67$ and the radial distribution function $g(r)$ obtained from simulation and DFT.
    }
    \label{fig:results}
\end{figure*}

We train a convolutional neural network to directly learn the excess free energy $\mathscr{F}_{\text{exc}}[\rho]$, enabling the straightforward calculation of functional derivatives by (auto-)differentiating the neural functional with respect to its inputs. We focus on 3D systems in a planar geometry, where the excess free-energy of a system of area $A$ is a functional of the density $\rho(z)$, which is constant across any plane parallel to the $xy$-plane, i.e., $\rho(z) = \rho(x,y,z)$ with $\rho(x, y, z) = \rho(x', y', z)$ for all $(x, y), (x', y')$ within the confines of $A$. We represent the excess free-energy functional $\mathscr{F}_{\text{exc}}[\rho]$ as a neural network $F_\theta^{(2)}(\rho_1,\cdots\rho_n)$, with $\rho_i$ the density at grid point $z_i$ for $i \in \{1, \ldots, n\}$, network parameters $\theta$, and where the upper index `$(2)$' indicates that the neural network is optimized for pair correlations of bulk systems obtained from simulations. 

The functional derivative $\delta \mathscr{F}_{\text{exc}}/\delta \rho(z_i)$ is defined as the limit of the partial derivative  $\lim_{\Delta z \rightarrow 0} (1/\Delta z) \partial \mathscr{F}_{\text{exc}}/\partial \rho_i$ with $\rho_i = \rho(z_i)$ and $\Delta z$ the (uniform) grid spacing. We leverage this relationship by employing automatic differentiation (autodiff) \cite{baydinAutomaticDifferentiationMachine, paszkeAutomaticDifferentiationPyTorch} to approximate the first and second functional derivatives of $\mathscr{F}_{\text{exc}}[\rho]$ on a finite grid by $(1/\Delta z) \partial F_{\theta}^{(2)}/\partial \rho_i$ and $(1/\Delta z)^2 \partial^2 F_{\theta}^{(2)}/\partial \rho_i \partial \rho_j$, respectively. Here we note that the second functional derivative of $\mathscr{F}_{\text{exc}}[\rho]$ in planar geometry is related to the laterally integrated direct correlation function $\bar{c}_b^{(2)}(|z_i-z_j|)$ through 
\begin{equation}
    \frac{-\beta}{A}\frac{\delta^2 \mathscr{F}_{\text{exc}}[\rho]}{\delta \rho(z_i) \delta \rho(z_j)} =    \bar{c}_b^{(2)}(|z_i-z_j|)\equiv \int_{|z_i-z_j|}^{\infty} \hspace{-6mm}dr \: 2\pi r \: c_b^{(2)}(r),
\label{eq:c2_1}
\end{equation}
where the direct correlation function $c^{(2)}_b(r)$ of homogeneous bulk fluids at density $\rho_b$ are obtained from simulated radial distribution functions $g(r)$ through the bulk Ornstein-Zernike equation
\begin{equation}
    c^{(2)}_b(r) = \frac{1}{2\pi^2}\int_0^\infty \frac{\sin(kr)}{kr}\left(\frac{\hat{h}(k)}{1 + \rho_b \hat{h}(k)}\right)k^2dk,
\label{eq:c2_2}
\end{equation}
with $\hat{h}(k)$ the Fourier transform of the total correlation function $h(r)=g(r)-1$ \cite{hansenTheorySimpleLiquids2013}. Thus we train our neural network such that its Hessian $\partial^2 F_{\theta}^{(2)}/\partial \rho_i \partial \rho_j$ represents well the laterally integrated direct correlation function obtained from homogeneous bulk simulations (see End Matter for further details). We will refer to this approach as pair-correlation matching.

We illustrate this methodology for systems interacting with a Lennard-Jones potential truncated at $r_{\text{cut}} = 4\sigma$ with $\sigma$ the particle diameter, shifted upwards by $\epsilon_{\text{cut}} = 0.98 \cdot 10^{-3} \epsilon$ with $\epsilon$ the well depth, at a temperature $k_BT/\epsilon = 2$, i.e.\ above the critical point. Setting $\Lambda=\sigma$ throughout, we perform grand-canonical Monte Carlo (GCMC) simulations of homogeneous bulk systems at a variety of chemical potentials $\beta\mu \in [-4, 0.5]$, resulting in bulk densities $\rho_b\sigma^3\in[0.02,0.67]$. We employ $10^9$ trial moves in a cubic box with an edge length $L=10\sigma$ (hence with area $A=L^2$ and volume $V=L^3$) and apply periodic boundary conditions. We sample $g(r)$ and $\rho_b$ and convert each $g(r)$ to $\bar{c}_b^{(2)}(z)$ using Eqs.~(\ref{eq:c2_1}) and (\ref{eq:c2_2}) with a grid-spacing $\Delta z=\sigma/32$. We create a data set of 800 combinations of $\bar{c}_b^{(2)}(z)$ and $\rho_b$, from which we learn the parameters $\theta$ of $F_\theta^{(2)}$. The neural network architecture of $F_\theta^{(2)}$ is a six-layer convolutional neural network with periodic padding \cite{alguacilEffectsBoundaryConditions2021} to match the system's periodic boundary conditions. Each convolution uses a kernel size of 3 with a dilation rate of 3. The number of channels per hidden layer is configured as follows: $[16,16,32,32,64,64]$, such that $\theta$ consists of 24.4K parameters (see End Matter for further details).

To evaluate the accuracy of our neural excess free-energy functional $F_\theta^{(2)}$, we compare it to the Van der Waals-like mean-field approximation $F_{\text{exc}}^{\text{MF}}$, which treats the attractions of Lennard-Jones particles as a perturbation on the hard-sphere system, as implemented in PyDFTlj  \cite{soaresClassicalDensityFunctional2023}. We use the White-Bear mark II version of FMT  for the excess free energy of the hard-sphere system \cite{hansen-goosDensityFunctionalTheory2006}. Additionally, we compare to $F_\theta^{(1)}$, which is a neural free-energy functional trained by minimization of the error between $(1/\Delta z)\partial F^{(1)}_{\theta}/\partial \rho_i$ and $\delta \mathscr{F}_{\text{exc}}/\delta \rho(z_i)$ rather than  by pair-correlation matching. This neural functional has the same neural network architecture as $F_\theta^{(2)}$ and is trained on a dataset of 800 non-uniform densities, subjected to the same set of chemical potentials as before (Supplementary Material, S.~2 \cite{SupplementaryMaterial}). Note that the free-energy functional $F_\theta^{(1)}$ introduced in this work is different from the neural direct correlation functional introduced by \textcite{sammullerNeuralFunctionalTheory2023a}, since the gradient of $F_\theta^{(1)}$ supplies a global estimate of $\delta \mathscr{F}_{\text{exc}}/\delta \rho(z_i)$ for all $i \in \{1,\dots,n\}$, in contrast to a local estimate at one position $z_i$ (Supplementary Material, S.~8 \cite{SupplementaryMaterial}). By approximating $\delta \mathscr{F}_{\text{exc}}/\delta \rho(z_i)$ by the gradient $(1/\Delta z)\partial F^{(n)}_{\theta}/\partial \rho_i$ for $n=1$ and 2, both neural functionals are applied in Picard iterations \cite{rothIntroductionDensityFunctional2006, edelmannNumericalEfficientWay2016, mairhoferNumericalAspectsClassical2017} to obtain DFT estimates for the equilibrium density profiles of inhomogeneous systems according to Eq.~(\ref{eq:equil_dens}).

The DFT results for an exemplary external potential at $\beta\mu=0$ are shown in Fig.~\ref{fig:results}a. We observe that the neural functionals $F_{\theta}^{(1)}$ and $F_{\theta}^{(2)}$ provide similar estimates that closely agree with simulation data. For the same external potential, we evaluate the accuracy of DFT estimates for the free energy for a range of chemical potentials $-4 < \beta \mu < 6$ (Fig.~\ref{fig:results}b). We compare with the excess free energy obtained from GCMC simulations through thermodynamic integration (Supplementary Material, S.~4 \cite{SupplementaryMaterial}). We observe that both neural functionals outperform $F_{\text{exc}}^{\text{MF}}$ within the range of $\mu$ values in the training set, exhibiting good agreement with simulations. The DFT estimates are shown until the Picard iterations diverge. We observe that $F_{\theta}^{(1)}$ diverges rapidly when extrapolating beyond the training set, even earlier than $F^{\text{exc}}_{\text{MF}}$. In contrast, $F_{\theta}^{(2)}$ converges to a solution far beyond the trained $\mu$ range.

For a more detailed comparison of the accuracy of the free-energy functionals for various inhomogeneous systems, we performed separate DFT calculations for 150 distinct external potentials, evaluated across the range $-4 < \beta \mu < 6$. Both $F_{\theta}^{(1)}$ and $F_{\theta}^{(2)}$ functionals show excellent agreement with simulated data for density estimates (Fig.~\ref{fig:results}c) and free-energy estimates (Fig.~\ref{fig:results}d), outperforming $F_{\text{exc}}^{\text{MF}}$ across all evaluated external potentials. Within the training range, $F_{\theta}^{(1)}$ exhibits marginally lower prediction errors than $F_{\theta}^{(2)}$, while $F_{\theta}^{(2)}$ demonstrates superior accuracy when extrapolating beyond the training range. Additional experiments show that the predictive accuracy of \(F_{\theta}^{(2)}\) is more sensitive to highly inhomogeneous densities compared to $F_{\theta}^{(1)}$, though this effect is reduced when incorporating higher bulk densities into the training set of $F_{\theta}^{(2)}$ (Supplementary Material, S.~7 \cite{SupplementaryMaterial}). Although we might expect that training on bulk data provides significantly less information about the underlying functional than training on inhomogeneous data, these results show that pair-correlation matching works remarkably well in practice.

Additionally, the free-energy functionals can be applied in a uniform density setting to obtain access to the bulk pressure 
\begin{equation}
    P(\rho_b) = \left(\frac{\delta\mathscr{F}_{\text{exc}}}{\delta \rho}\Big|_{\rho_b} + k_BT \right) \rho_b - \mathscr{F}_{\text{exc}}[\rho_b]/V, 
\end{equation}
and the chemical potential $\mu$ following Eq.~(\ref{eq:equil_dens}). Again, we find excellent agreement of both neural functionals with simulations within the training set, and superior agreement of $F_{\theta}^{(2)}$ at higher densities, as shown in Fig.~\ref{fig:results}e. Lastly, we demonstrate that $F_{\theta}^{(2)}$ compared to both $F_{\theta}^{(1)}$ and $F^{\text{exc}}_{\text{MF}}$ provides accurate estimates for the laterally integrated direct correlation function 
$\bar{c}_b^{(2)}(z)$ and the radial distribution function $g(r)$ as shown in Fig.~\ref{fig:results}f for $\rho_b\sigma^3 = 0.67$. 
Here, we approximate $\bar{c}_b^{(2)}(z)$ by $(-\beta/A(\Delta z)^2) \partial^2 F_{\theta}/\partial \rho_i \partial \rho_j$ with $z=|z_i-z_j|$. To derive the radial distribution function, we first numerically calculate $c_b^{(2)}(r)$ from $\bar{c}_b^{(2)}(z)$ using
\begin{equation}
     -\left.\frac{1}{2 \pi}\left(\frac{1}{z} \frac{d \bar{c}_b^{(2)}(z)}{d z}\right)\right|_{z=r}= c_b^{(2)}(r).
\end{equation}
We then obtain  $g(r)$  using the Ornstein-Zernike equation (Eq.~\ref{eq:c2_2}). To suppress artefacts stemming from numerical transformations in the region for $g(r) \rightarrow 0$, we apply a noise-reducing filter (Supplementary Material, S.~5 \cite{SupplementaryMaterial}).


Our results lead to the surprising observation that the neural free-energy functional $F_\theta^{(2)}$ is robust and predicts accurate non-uniform density profiles, even substantially beyond its training range and solely by training on pair correlation functions of homogeneous bulk systems. This training on homogeneous bulk systems offers an alternative to existing training schemes of classical DFT, which so far have all been based on training on density profiles of inhomogeneous fluids in a variety of external fields \cite{linClassicalDensityFunctional2019,catsMachinelearningFreeenergyFunctionals2021,sammullerNeuralFunctionalTheory2023a,simonMachineLearningDensity2024}. For pairwise systems, this alternative could be particularly useful if pair-correlations are the only data available (e.g. from scattering experiments on homogeneous fluids), if the external potential leads to computationally expensive 3D density profiles, or if spatial heterogeneity is coupled to orientational anisotropy (e.g.\ in liquid crystals, non-spherical molecules, or patchy particles \cite{simonMachineLearningDensity2024}). The presently proposed training on homogeneous bulk systems can also be a computationally inexpensive addition to the existing training protocols on inhomogeneous fluids, where an optimal combination may be found between training on homogeneous and inhomogeneous states. In this sense, the present study can be seen as an extreme case of training on pair correlations in the bulk only. 

In conclusion, this work introduces a generic machine learning approach to obtain  classical free-energy functionals through pair-correlation matching, through which we attain a neural free-energy functional that enables simultaneous and direct access to both the excess free energy and the density of a classical supercritical system of interacting particles in any external environment.


\begin{acknowledgments}
\section*{Acknowledgements} 
The authors acknowledge the University of Amsterdam Data Science Centre for financial support. M.D. acknowledges funding from the European
Research Council (ERC) under the European Union’s Horizon
2020 Research and Innovation Program (Grant Agreement No.
ERC-2019-ADG 884902 SoftML). J.W.M. acknowledges funding from the European Union's Horizon Framework Programme (Grant agreement ID: 101120237). 
\end{acknowledgments}


\nocite{sammullerNeuralDensityFunctionals2024} 

\bibliography{references}

\begin{thebibliography}{23}%
\makeatletter
\providecommand \@ifxundefined [1]{%
 \@ifx{#1\undefined}
}%
\providecommand \@ifnum [1]{%
 \ifnum #1\expandafter \@firstoftwo
 \else \expandafter \@secondoftwo
 \fi
}%
\providecommand \@ifx [1]{%
 \ifx #1\expandafter \@firstoftwo
 \else \expandafter \@secondoftwo
 \fi
}%
\providecommand \natexlab [1]{#1}%
\providecommand \enquote  [1]{``#1''}%
\providecommand \bibnamefont  [1]{#1}%
\providecommand \bibfnamefont [1]{#1}%
\providecommand \citenamefont [1]{#1}%
\providecommand \href@noop [0]{\@secondoftwo}%
\providecommand \href [0]{\begingroup \@sanitize@url \@href}%
\providecommand \@href[1]{\@@startlink{#1}\@@href}%
\providecommand \@@href[1]{\endgroup#1\@@endlink}%
\providecommand \@sanitize@url [0]{\catcode `\\12\catcode `\$12\catcode
  `\&12\catcode `\#12\catcode `\^12\catcode `\_12\catcode `\%12\relax}%
\providecommand \@@startlink[1]{}%
\providecommand \@@endlink[0]{}%
\providecommand \url  [0]{\begingroup\@sanitize@url \@url }%
\providecommand \@url [1]{\endgroup\@href {#1}{\urlprefix }}%
\providecommand \urlprefix  [0]{URL }%
\providecommand \Eprint [0]{\href }%
\providecommand \doibase [0]{https://doi.org/}%
\providecommand \selectlanguage [0]{\@gobble}%
\providecommand \bibinfo  [0]{\@secondoftwo}%
\providecommand \bibfield  [0]{\@secondoftwo}%
\providecommand \translation [1]{[#1]}%
\providecommand \BibitemOpen [0]{}%
\providecommand \bibitemStop [0]{}%
\providecommand \bibitemNoStop [0]{.\EOS\space}%
\providecommand \EOS [0]{\spacefactor3000\relax}%
\providecommand \BibitemShut  [1]{\csname bibitem#1\endcsname}%
\let\auto@bib@innerbib\@empty
\bibitem [{\citenamefont {Soares}\ \emph {et~al.}(2023)\citenamefont {Soares},
  \citenamefont {Barreto~Jr.},\ and\ \citenamefont
  {Tavares}}]{soaresClassicalDensityFunctional2023}%
  \BibitemOpen
  \bibfield  {author} {\bibinfo {author} {\bibfnamefont {E.~d.~A.}\
  \bibnamefont {Soares}}, \bibinfo {author} {\bibfnamefont {A.~G.}\
  \bibnamefont {Barreto~Jr.}},\ and\ \bibinfo {author} {\bibfnamefont {F.~W.}\
  \bibnamefont {Tavares}},\ }\href@noop {} {\bibinfo {title} {Classical
  {{Density Functional Theory Reveals Structural Information}} of {{H2}} and
  {{CH4 Fluids Adsorbed}} in {{MOF-5}}}} (\bibinfo {year} {2023}),\ \Eprint
  {https://arxiv.org/abs/2303.11384} {arxiv:2303.11384 [cond-mat,
  physics:physics]} \BibitemShut {NoStop}%
\bibitem [{\citenamefont {Guo}\ \emph {et~al.}(2016)\citenamefont {Guo},
  \citenamefont {Liu}, \citenamefont {Hu}, \citenamefont {Liu},\ and\
  \citenamefont {Hu}}]{guoClassicalDensityFunctional2016}%
  \BibitemOpen
  \bibfield  {author} {\bibinfo {author} {\bibfnamefont {F.}~\bibnamefont
  {Guo}}, \bibinfo {author} {\bibfnamefont {Y.}~\bibnamefont {Liu}}, \bibinfo
  {author} {\bibfnamefont {J.}~\bibnamefont {Hu}}, \bibinfo {author}
  {\bibfnamefont {H.}~\bibnamefont {Liu}},\ and\ \bibinfo {author}
  {\bibfnamefont {Y.}~\bibnamefont {Hu}},\ }\bibfield  {title} {\bibinfo
  {title} {Classical density functional theory for gas separation in nanoporous
  materials and its application to {{CH4}}/{{H2}} separation},\ }\href
  {https://doi.org/10.1016/j.ces.2016.04.027} {\bibfield  {journal} {\bibinfo
  {journal} {Chem.\ Eng.\ Sci.}\ }\textbf {\bibinfo {volume} {149}},\ \bibinfo
  {pages} {14} (\bibinfo {year} {2016})}\BibitemShut {NoStop}%
\bibitem [{\citenamefont {Guo}\ \emph {et~al.}(2018)\citenamefont {Guo},
  \citenamefont {Liu}, \citenamefont {Hu}, \citenamefont {Liu},\ and\
  \citenamefont {Hu}}]{guoFastScreeningPorous2018}%
  \BibitemOpen
  \bibfield  {author} {\bibinfo {author} {\bibfnamefont {F.}~\bibnamefont
  {Guo}}, \bibinfo {author} {\bibfnamefont {Y.}~\bibnamefont {Liu}}, \bibinfo
  {author} {\bibfnamefont {J.}~\bibnamefont {Hu}}, \bibinfo {author}
  {\bibfnamefont {H.}~\bibnamefont {Liu}},\ and\ \bibinfo {author}
  {\bibfnamefont {Y.}~\bibnamefont {Hu}},\ }\bibfield  {title} {\bibinfo
  {title} {Fast screening of porous materials for noble gas adsorption and
  separation: A classical density functional approach},\ }\href
  {https://doi.org/10.1039/C8CP03777A} {\bibfield  {journal} {\bibinfo
  {journal} {Phys.\ Chem.\ Chem.\ Phys.}\ }\textbf {\bibinfo {volume} {20}},\
  \bibinfo {pages} {28193} (\bibinfo {year} {2018})}\BibitemShut {NoStop}%
\bibitem [{\citenamefont {Guo}\ \emph {et~al.}(2020)\citenamefont {Guo},
  \citenamefont {Liu}, \citenamefont {Hu}, \citenamefont {Liu},\ and\
  \citenamefont {Hu}}]{guoScreeningPorousMaterials2020}%
  \BibitemOpen
  \bibfield  {author} {\bibinfo {author} {\bibfnamefont {F.}~\bibnamefont
  {Guo}}, \bibinfo {author} {\bibfnamefont {Y.}~\bibnamefont {Liu}}, \bibinfo
  {author} {\bibfnamefont {J.}~\bibnamefont {Hu}}, \bibinfo {author}
  {\bibfnamefont {H.}~\bibnamefont {Liu}},\ and\ \bibinfo {author}
  {\bibfnamefont {Y.}~\bibnamefont {Hu}},\ }\bibfield  {title} {\bibinfo
  {title} {Screening of {{Porous Materials}} for {{Toxic Gas Adsorption}}:
  {{Classical Density Functional Approach}}},\ }\href
  {https://doi.org/10.1021/acs.iecr.0c02659} {\bibfield  {journal} {\bibinfo
  {journal} {Ind.\ Eng.\ Chem.\ Res.}\ }\textbf {\bibinfo {volume} {59}},\
  \bibinfo {pages} {14364} (\bibinfo {year} {2020})}\BibitemShut {NoStop}%
\bibitem [{\citenamefont {Fu}\ \emph {et~al.}(2015)\citenamefont {Fu},
  \citenamefont {Tian},\ and\ \citenamefont
  {Wu}}]{fuClassicalDensityFunctional2015}%
  \BibitemOpen
  \bibfield  {author} {\bibinfo {author} {\bibfnamefont {J.}~\bibnamefont
  {Fu}}, \bibinfo {author} {\bibfnamefont {Y.}~\bibnamefont {Tian}},\ and\
  \bibinfo {author} {\bibfnamefont {J.}~\bibnamefont {Wu}},\ }\bibfield
  {title} {\bibinfo {title} {Classical density functional theory for methane
  adsorption in metal-organic framework materials},\ }\href
  {https://doi.org/10.1002/aic.14877} {\bibfield  {journal} {\bibinfo
  {journal} {AIChE J.}\ }\textbf {\bibinfo {volume} {61}},\ \bibinfo {pages}
  {3012} (\bibinfo {year} {2015})}\BibitemShut {NoStop}%
\bibitem [{\citenamefont {Liu}\ and\ \citenamefont
  {Liu}(2016)}]{liuClassicalDensityFunctional2016}%
  \BibitemOpen
  \bibfield  {author} {\bibinfo {author} {\bibfnamefont {Y.}~\bibnamefont
  {Liu}}\ and\ \bibinfo {author} {\bibfnamefont {H.}~\bibnamefont {Liu}},\
  }\bibfield  {title} {\bibinfo {title} {Classical {{Density Functional
  Theory}} for {{Fluids Adsorption}} in {{MOFs}}},\ }in\ \href
  {https://doi.org/10.5772/64632} {\emph {\bibinfo {booktitle} {Metal-{{Organic
  Frameworks}}}}},\ \bibinfo {editor} {edited by\ \bibinfo {editor}
  {\bibfnamefont {F.}~\bibnamefont {Zafar}}\ and\ \bibinfo {editor}
  {\bibfnamefont {E.}~\bibnamefont {Sharmin}}}\ (\bibinfo  {publisher}
  {InTech},\ \bibinfo {year} {2016})\BibitemShut {NoStop}%
\bibitem [{\citenamefont {Evans}(1979)}]{evansNatureLiquidvapourInterface1979}%
  \BibitemOpen
  \bibfield  {author} {\bibinfo {author} {\bibfnamefont {R.}~\bibnamefont
  {Evans}},\ }\bibfield  {title} {\bibinfo {title} {The nature of the
  liquid-vapour interface and other topics in the statistical mechanics of
  non-uniform, classical fluids},\ }\href
  {https://doi.org/10.1080/00018737900101365} {\bibfield  {journal} {\bibinfo
  {journal} {Adv.\ Phys.}\ }\textbf {\bibinfo {volume} {28}},\ \bibinfo {pages}
  {143} (\bibinfo {year} {1979})}\BibitemShut {NoStop}%
\bibitem [{\citenamefont {Hansen}\ and\ \citenamefont
  {McDonald}(2013)}]{hansenTheorySimpleLiquids2013}%
  \BibitemOpen
  \bibfield  {author} {\bibinfo {author} {\bibfnamefont {J.-P.}\ \bibnamefont
  {Hansen}}\ and\ \bibinfo {author} {\bibfnamefont {I.~R.}\ \bibnamefont
  {McDonald}},\ }\bibfield  {title} {\bibinfo {title} {Theory of {{Simple
  Liquids}}},\ }in\ \href {https://doi.org/10.1016/B978-0-12-387032-2.00001-5}
  {\emph {\bibinfo {booktitle} {Theory of {{Simple Liquids}}}}}\ (\bibinfo
  {publisher} {Elsevier},\ \bibinfo {year} {2013})\BibitemShut {NoStop}%
\bibitem [{\citenamefont {Roth}(2010)}]{rothFundamentalMeasureTheory2010}%
  \BibitemOpen
  \bibfield  {author} {\bibinfo {author} {\bibfnamefont {R.}~\bibnamefont
  {Roth}},\ }\bibfield  {title} {\bibinfo {title} {Fundamental measure theory
  for hard-sphere mixtures: {{A}} review},\ }\bibfield  {journal} {\bibinfo
  {journal} {J.\ Phys.\ Condens.\ Matter}\ }\textbf {\bibinfo {volume} {22}},\
  \href {https://doi.org/10.1088/0953-8984/22/6/063102}
  {10.1088/0953-8984/22/6/063102} (\bibinfo {year} {2010})\BibitemShut
  {NoStop}%
\bibitem [{\citenamefont {Cats}\ \emph
  {et~al.}(2021{\natexlab{a}})\citenamefont {Cats}, \citenamefont {Evans},
  \citenamefont {H{\"a}rtel},\ and\ \citenamefont
  {Van~Roij}}]{catsPrimitiveModelElectrolytes2021}%
  \BibitemOpen
  \bibfield  {author} {\bibinfo {author} {\bibfnamefont {P.}~\bibnamefont
  {Cats}}, \bibinfo {author} {\bibfnamefont {R.}~\bibnamefont {Evans}},
  \bibinfo {author} {\bibfnamefont {A.}~\bibnamefont {H{\"a}rtel}},\ and\
  \bibinfo {author} {\bibfnamefont {R.}~\bibnamefont {Van~Roij}},\ }\bibfield
  {title} {\bibinfo {title} {Primitive model electrolytes in the near and far
  field: {{Decay}} lengths from {{DFT}} and simulations},\ }\bibfield
  {journal} {\bibinfo  {journal} {J.\ Chem.\ Phys.}\ }\textbf {\bibinfo
  {volume} {154}},\ \href {https://doi.org/10.1063/5.0039619}
  {10.1063/5.0039619} (\bibinfo {year} {2021}{\natexlab{a}}),\ \Eprint
  {https://arxiv.org/abs/2012.02713} {arxiv:2012.02713} \BibitemShut {NoStop}%
\bibitem [{\citenamefont {Lin}\ and\ \citenamefont
  {Oettel}(2019)}]{linClassicalDensityFunctional2019}%
  \BibitemOpen
  \bibfield  {author} {\bibinfo {author} {\bibfnamefont {S.~C.}\ \bibnamefont
  {Lin}}\ and\ \bibinfo {author} {\bibfnamefont {M.}~\bibnamefont {Oettel}},\
  }\bibfield  {title} {\bibinfo {title} {A classical density functional from
  machine learning and a convolutional neural network},\ }\href
  {https://doi.org/10.21468/SciPostPhys.6.2.025} {\bibfield  {journal}
  {\bibinfo  {journal} {SciPost Phys.}\ }\textbf {\bibinfo {volume} {6}},\
  \bibinfo {pages} {1} (\bibinfo {year} {2019})},\ \Eprint
  {https://arxiv.org/abs/1811.05728} {arxiv:1811.05728} \BibitemShut {NoStop}%
\bibitem [{\citenamefont {Cats}\ \emph
  {et~al.}(2021{\natexlab{b}})\citenamefont {Cats}, \citenamefont {Kuipers},
  \citenamefont {De~Wind}, \citenamefont {Van~Damme}, \citenamefont {Coli},
  \citenamefont {Dijkstra},\ and\ \citenamefont
  {Van~Roij}}]{catsMachinelearningFreeenergyFunctionals2021}%
  \BibitemOpen
  \bibfield  {author} {\bibinfo {author} {\bibfnamefont {P.}~\bibnamefont
  {Cats}}, \bibinfo {author} {\bibfnamefont {S.}~\bibnamefont {Kuipers}},
  \bibinfo {author} {\bibfnamefont {S.}~\bibnamefont {De~Wind}}, \bibinfo
  {author} {\bibfnamefont {R.}~\bibnamefont {Van~Damme}}, \bibinfo {author}
  {\bibfnamefont {G.~M.}\ \bibnamefont {Coli}}, \bibinfo {author}
  {\bibfnamefont {M.}~\bibnamefont {Dijkstra}},\ and\ \bibinfo {author}
  {\bibfnamefont {R.}~\bibnamefont {Van~Roij}},\ }\bibfield  {title} {\bibinfo
  {title} {Machine-learning free-energy functionals using density profiles from
  simulations},\ }\bibfield  {journal} {\bibinfo  {journal} {APL Mater.}\
  }\textbf {\bibinfo {volume} {9}},\ \href {https://doi.org/10.1063/5.0042558}
  {10.1063/5.0042558} (\bibinfo {year} {2021}{\natexlab{b}}),\ \Eprint
  {https://arxiv.org/abs/2101.01942} {arxiv:2101.01942} \BibitemShut {NoStop}%
\bibitem [{\citenamefont {Samm{\"u}ller}\ \emph {et~al.}(2023)\citenamefont
  {Samm{\"u}ller}, \citenamefont {Hermann}, \citenamefont {De~Las~Heras},\ and\
  \citenamefont {Schmidt}}]{sammullerNeuralFunctionalTheory2023a}%
  \BibitemOpen
  \bibfield  {author} {\bibinfo {author} {\bibfnamefont {F.}~\bibnamefont
  {Samm{\"u}ller}}, \bibinfo {author} {\bibfnamefont {S.}~\bibnamefont
  {Hermann}}, \bibinfo {author} {\bibfnamefont {D.}~\bibnamefont
  {De~Las~Heras}},\ and\ \bibinfo {author} {\bibfnamefont {M.}~\bibnamefont
  {Schmidt}},\ }\bibfield  {title} {\bibinfo {title} {Neural functional theory
  for inhomogeneous fluids: {{Fundamentals}} and applications},\ }\href
  {https://doi.org/10.1073/pnas.2312484120} {\bibfield  {journal} {\bibinfo
  {journal} {Proc.\ Natl.\ Acad.\ Sci. USA}\ }\textbf {\bibinfo {volume}
  {120}},\ \bibinfo {pages} {e2312484120} (\bibinfo {year} {2023})}\BibitemShut
  {NoStop}%
\bibitem [{\citenamefont {Simon}\ \emph {et~al.}(2024)\citenamefont {Simon},
  \citenamefont {Weimar}, \citenamefont {Martius},\ and\ \citenamefont
  {Oettel}}]{simonMachineLearningDensity2024}%
  \BibitemOpen
  \bibfield  {author} {\bibinfo {author} {\bibfnamefont {A.}~\bibnamefont
  {Simon}}, \bibinfo {author} {\bibfnamefont {J.}~\bibnamefont {Weimar}},
  \bibinfo {author} {\bibfnamefont {G.}~\bibnamefont {Martius}},\ and\ \bibinfo
  {author} {\bibfnamefont {M.}~\bibnamefont {Oettel}},\ }\bibfield  {title}
  {\bibinfo {title} {Machine {{Learning}} of a {{Density Functional}} for
  {{Anisotropic Patchy Particles}}},\ }\href
  {https://doi.org/10.1021/acs.jctc.3c01238} {\bibfield  {journal} {\bibinfo
  {journal} {J.\ Chem.\ Theory Comput.}\ }\textbf {\bibinfo {volume} {20}},\
  \bibinfo {pages} {1062} (\bibinfo {year} {2024})}\BibitemShut {NoStop}%
\bibitem [{\citenamefont {Roth}(2006)}]{rothIntroductionDensityFunctional2006}%
  \BibitemOpen
  \bibfield  {author} {\bibinfo {author} {\bibfnamefont {R.}~\bibnamefont
  {Roth}},\ }\bibfield  {title} {\bibinfo {title} {Introduction to {{Density
  Functional Theory}} of {{Classical Systems}}: {{Theory}} and
  {{Applications}}},\ }\href@noop {} {\bibfield  {journal} {\bibinfo  {journal}
  {Lecture Notes}\ } (\bibinfo {year} {2006})}\BibitemShut {NoStop}%
\bibitem [{\citenamefont {Edelmann}\ and\ \citenamefont
  {Roth}(2016)}]{edelmannNumericalEfficientWay2016}%
  \BibitemOpen
  \bibfield  {author} {\bibinfo {author} {\bibfnamefont {M.}~\bibnamefont
  {Edelmann}}\ and\ \bibinfo {author} {\bibfnamefont {R.}~\bibnamefont
  {Roth}},\ }\bibfield  {title} {\bibinfo {title} {A numerical efficient way to
  minimize classical density functional theory},\ }\href
  {https://doi.org/10.1063/1.4942020} {\bibfield  {journal} {\bibinfo
  {journal} {J.\ Chem.\ Phys.}\ }\textbf {\bibinfo {volume} {144}},\ \bibinfo
  {pages} {074105} (\bibinfo {year} {2016})}\BibitemShut {NoStop}%
\bibitem [{\citenamefont {Mairhofer}\ and\ \citenamefont
  {Gross}(2017)}]{mairhoferNumericalAspectsClassical2017}%
  \BibitemOpen
  \bibfield  {author} {\bibinfo {author} {\bibfnamefont {J.}~\bibnamefont
  {Mairhofer}}\ and\ \bibinfo {author} {\bibfnamefont {J.}~\bibnamefont
  {Gross}},\ }\bibfield  {title} {\bibinfo {title} {Numerical aspects of
  classical density functional theory for one-dimensional vapor-liquid
  interfaces},\ }\href {https://doi.org/10.1016/j.fluid.2017.03.023} {\bibfield
   {journal} {\bibinfo  {journal} {Fluid Phase Equilibria}\ }\textbf {\bibinfo
  {volume} {444}},\ \bibinfo {pages} {1} (\bibinfo {year} {2017})}\BibitemShut
  {NoStop}%
\bibitem [{\citenamefont {Baydin}\ \emph {et~al.}(2018)\citenamefont {Baydin},
  \citenamefont {Pearlmutter}, \citenamefont {Radul},\ and\ \citenamefont
  {Siskind}}]{baydinAutomaticDifferentiationMachine}%
  \BibitemOpen
  \bibfield  {author} {\bibinfo {author} {\bibfnamefont {A.~G.}\ \bibnamefont
  {Baydin}}, \bibinfo {author} {\bibfnamefont {B.~A.}\ \bibnamefont
  {Pearlmutter}}, \bibinfo {author} {\bibfnamefont {A.~A.}\ \bibnamefont
  {Radul}},\ and\ \bibinfo {author} {\bibfnamefont {J.~M.}\ \bibnamefont
  {Siskind}},\ }\bibfield  {title} {\bibinfo {title} {Automatic differentiation
  in machine learning: a survey},\ }\href
  {http://jmlr.org/papers/v18/17-468.html} {\bibfield  {journal} {\bibinfo
  {journal} {J. Machine Learn. Res.}\ }\textbf {\bibinfo {volume} {18}},\
  \bibinfo {pages} {1} (\bibinfo {year} {2018})}\BibitemShut {NoStop}%
\bibitem [{\citenamefont {Paszke}\ \emph {et~al.}(2017)\citenamefont {Paszke},
  \citenamefont {Gross}, \citenamefont {Chintala}, \citenamefont {Chanan},
  \citenamefont {Yang}, \citenamefont {DeVito}, \citenamefont {Lin},
  \citenamefont {Desmaison}, \citenamefont {Antiga},\ and\ \citenamefont
  {Lerer}}]{paszkeAutomaticDifferentiationPyTorch}%
  \BibitemOpen
  \bibfield  {author} {\bibinfo {author} {\bibfnamefont {A.}~\bibnamefont
  {Paszke}}, \bibinfo {author} {\bibfnamefont {S.}~\bibnamefont {Gross}},
  \bibinfo {author} {\bibfnamefont {S.}~\bibnamefont {Chintala}}, \bibinfo
  {author} {\bibfnamefont {G.}~\bibnamefont {Chanan}}, \bibinfo {author}
  {\bibfnamefont {E.}~\bibnamefont {Yang}}, \bibinfo {author} {\bibfnamefont
  {Z.}~\bibnamefont {DeVito}}, \bibinfo {author} {\bibfnamefont
  {Z.}~\bibnamefont {Lin}}, \bibinfo {author} {\bibfnamefont {A.}~\bibnamefont
  {Desmaison}}, \bibinfo {author} {\bibfnamefont {L.}~\bibnamefont {Antiga}},\
  and\ \bibinfo {author} {\bibfnamefont {A.}~\bibnamefont {Lerer}},\ }\bibfield
   {title} {\bibinfo {title} {Automatic differentiation in {{PyTorch}}},\
  }\href@noop {} {\bibfield  {journal} {\bibinfo  {journal} {NIPS Autodiff
  Workshop}\ } (\bibinfo {year} {2017})}\BibitemShut {NoStop}%
\bibitem [{\citenamefont {Alguacil}\ \emph {et~al.}(2021)\citenamefont
  {Alguacil}, \citenamefont {Pinto}, \citenamefont {Bauerheim}, \citenamefont
  {Jacob},\ and\ \citenamefont
  {Moreau}}]{alguacilEffectsBoundaryConditions2021}%
  \BibitemOpen
  \bibfield  {author} {\bibinfo {author} {\bibfnamefont {A.}~\bibnamefont
  {Alguacil}}, \bibinfo {author} {\bibfnamefont {W.~G.}\ \bibnamefont {Pinto}},
  \bibinfo {author} {\bibfnamefont {M.}~\bibnamefont {Bauerheim}}, \bibinfo
  {author} {\bibfnamefont {M.~C.}\ \bibnamefont {Jacob}},\ and\ \bibinfo
  {author} {\bibfnamefont {S.}~\bibnamefont {Moreau}},\ }\href@noop {}
  {\bibinfo {title} {Effects of boundary conditions in fully convolutional
  networks for learning spatio-temporal dynamics}} (\bibinfo {year} {2021}),\
  \Eprint {https://arxiv.org/abs/2106.11160} {arxiv:2106.11160 [physics]}
  \BibitemShut {NoStop}%
\bibitem [{\citenamefont {{Hansen-Goos}}\ and\ \citenamefont
  {Roth}(2006)}]{hansen-goosDensityFunctionalTheory2006}%
  \BibitemOpen
  \bibfield  {author} {\bibinfo {author} {\bibfnamefont {H.}~\bibnamefont
  {{Hansen-Goos}}}\ and\ \bibinfo {author} {\bibfnamefont {R.}~\bibnamefont
  {Roth}},\ }\bibfield  {title} {\bibinfo {title} {Density functional theory
  for hard-sphere mixtures: The {{White Bear}} version mark {{II}}},\ }\href
  {https://doi.org/10.1088/0953-8984/18/37/002} {\bibfield  {journal} {\bibinfo
   {journal} {J.\ Phys.\ Condens.\ Matter}\ }\textbf {\bibinfo {volume} {18}},\
  \bibinfo {pages} {8413} (\bibinfo {year} {2006})}\BibitemShut {NoStop}%
\bibitem [{Sup()}]{SupplementaryMaterial}%
  \BibitemOpen
  \href@noop {} {}\bibinfo {note} {See {{Supplementary Material}} at [URL
  Placeholder] for details on the presented methods and additional results,
  with references to
  {Ref.~\cite{sammullerNeuralDensityFunctionals2024}}.}\BibitemShut {Stop}%
\bibitem [{\citenamefont {Samm{\"u}ller}\ and\ \citenamefont
  {Schmidt}(2024)}]{sammullerNeuralDensityFunctionals2024}%
  \BibitemOpen
  \bibfield  {author} {\bibinfo {author} {\bibfnamefont {F.}~\bibnamefont
  {Samm{\"u}ller}}\ and\ \bibinfo {author} {\bibfnamefont {M.}~\bibnamefont
  {Schmidt}},\ }\bibfield  {title} {\bibinfo {title} {Neural density
  functionals: {{Local}} learning and pair-correlation matching},\ }\href
  {https://doi.org/10.1103/PhysRevE.110.L032601} {\bibfield  {journal}
  {\bibinfo  {journal} {Phys.\ Rev.\ E}\ }\textbf {\bibinfo {volume} {110}},\
  \bibinfo {pages} {L032601} (\bibinfo {year} {2024})}\BibitemShut {NoStop}%
\end{thebibliography}%


\begin{thebibliography}{8}%
\makeatletter
\providecommand \@ifxundefined [1]{%
 \@ifx{#1\undefined}
}%
\providecommand \@ifnum [1]{%
 \ifnum #1\expandafter \@firstoftwo
 \else \expandafter \@secondoftwo
 \fi
}%
\providecommand \@ifx [1]{%
 \ifx #1\expandafter \@firstoftwo
 \else \expandafter \@secondoftwo
 \fi
}%
\providecommand \natexlab [1]{#1}%
\providecommand \enquote  [1]{``#1''}%
\providecommand \bibnamefont  [1]{#1}%
\providecommand \bibfnamefont [1]{#1}%
\providecommand \citenamefont [1]{#1}%
\providecommand \href@noop [0]{\@secondoftwo}%
\providecommand \href [0]{\begingroup \@sanitize@url \@href}%
\providecommand \@href[1]{\@@startlink{#1}\@@href}%
\providecommand \@@href[1]{\endgroup#1\@@endlink}%
\providecommand \@sanitize@url [0]{\catcode `\\12\catcode `\$12\catcode
  `\&12\catcode `\#12\catcode `\^12\catcode `\_12\catcode `\%12\relax}%
\providecommand \@@startlink[1]{}%
\providecommand \@@endlink[0]{}%
\providecommand \url  [0]{\begingroup\@sanitize@url \@url }%
\providecommand \@url [1]{\endgroup\@href {#1}{\urlprefix }}%
\providecommand \urlprefix  [0]{URL }%
\providecommand \Eprint [0]{\href }%
\providecommand \doibase [0]{https://doi.org/}%
\providecommand \selectlanguage [0]{\@gobble}%
\providecommand \bibinfo  [0]{\@secondoftwo}%
\providecommand \bibfield  [0]{\@secondoftwo}%
\providecommand \translation [1]{[#1]}%
\providecommand \BibitemOpen [0]{}%
\providecommand \bibitemStop [0]{}%
\providecommand \bibitemNoStop [0]{.\EOS\space}%
\providecommand \EOS [0]{\spacefactor3000\relax}%
\providecommand \BibitemShut  [1]{\csname bibitem#1\endcsname}%
\let\auto@bib@innerbib\@empty
\bibitem [{\citenamefont {Evans}(1979)}]{evansNatureLiquidvapourInterface1979}%
  \BibitemOpen
  \bibfield  {author} {\bibinfo {author} {\bibfnamefont {R.}~\bibnamefont
  {Evans}},\ }\bibfield  {title} {\bibinfo {title} {The nature of the
  liquid-vapour interface and other topics in the statistical mechanics of
  non-uniform, classical fluids},\ }\href
  {https://doi.org/10.1080/00018737900101365} {\bibfield  {journal} {\bibinfo
  {journal} {Adv.\ Phys.}\ }\textbf {\bibinfo {volume} {28}},\ \bibinfo {pages}
  {143} (\bibinfo {year} {1979})}\BibitemShut {NoStop}%
\bibitem [{\citenamefont {Roth}(2006)}]{rothIntroductionDensityFunctional2006}%
  \BibitemOpen
  \bibfield  {author} {\bibinfo {author} {\bibfnamefont {R.}~\bibnamefont
  {Roth}},\ }\bibfield  {title} {\bibinfo {title} {Introduction to {{Density
  Functional Theory}} of {{Classical Systems}}: {{Theory}} and
  {{Applications}}},\ }\href@noop {} {\bibfield  {journal} {\bibinfo  {journal}
  {Lecture Notes}\ } (\bibinfo {year} {2006})}\BibitemShut {NoStop}%
\bibitem [{\citenamefont {Edelmann}\ and\ \citenamefont
  {Roth}(2016)}]{edelmannNumericalEfficientWay2016}%
  \BibitemOpen
  \bibfield  {author} {\bibinfo {author} {\bibfnamefont {M.}~\bibnamefont
  {Edelmann}}\ and\ \bibinfo {author} {\bibfnamefont {R.}~\bibnamefont
  {Roth}},\ }\bibfield  {title} {\bibinfo {title} {A numerical efficient way to
  minimize classical density functional theory},\ }\href
  {https://doi.org/10.1063/1.4942020} {\bibfield  {journal} {\bibinfo
  {journal} {J.\ Chem.\ Phys.}\ }\textbf {\bibinfo {volume} {144}},\ \bibinfo
  {pages} {074105} (\bibinfo {year} {2016})}\BibitemShut {NoStop}%
\bibitem [{\citenamefont {Mairhofer}\ and\ \citenamefont
  {Gross}(2017)}]{mairhoferNumericalAspectsClassical2017}%
  \BibitemOpen
  \bibfield  {author} {\bibinfo {author} {\bibfnamefont {J.}~\bibnamefont
  {Mairhofer}}\ and\ \bibinfo {author} {\bibfnamefont {J.}~\bibnamefont
  {Gross}},\ }\bibfield  {title} {\bibinfo {title} {Numerical aspects of
  classical density functional theory for one-dimensional vapor-liquid
  interfaces},\ }\href {https://doi.org/10.1016/j.fluid.2017.03.023} {\bibfield
   {journal} {\bibinfo  {journal} {Fluid Phase Equilibria}\ }\textbf {\bibinfo
  {volume} {444}},\ \bibinfo {pages} {1} (\bibinfo {year} {2017})}\BibitemShut
  {NoStop}%
\bibitem [{\citenamefont {Samm{\"u}ller}\ \emph {et~al.}(2023)\citenamefont
  {Samm{\"u}ller}, \citenamefont {Hermann}, \citenamefont {De~Las~Heras},\ and\
  \citenamefont {Schmidt}}]{sammullerNeuralFunctionalTheory2023a}%
  \BibitemOpen
  \bibfield  {author} {\bibinfo {author} {\bibfnamefont {F.}~\bibnamefont
  {Samm{\"u}ller}}, \bibinfo {author} {\bibfnamefont {S.}~\bibnamefont
  {Hermann}}, \bibinfo {author} {\bibfnamefont {D.}~\bibnamefont
  {De~Las~Heras}},\ and\ \bibinfo {author} {\bibfnamefont {M.}~\bibnamefont
  {Schmidt}},\ }\bibfield  {title} {\bibinfo {title} {Neural functional theory
  for inhomogeneous fluids: {{Fundamentals}} and applications},\ }\href
  {https://doi.org/10.1073/pnas.2312484120} {\bibfield  {journal} {\bibinfo
  {journal} {Proc.\ Natl.\ Acad.\ Sci. USA}\ }\textbf {\bibinfo {volume}
  {120}},\ \bibinfo {pages} {e2312484120} (\bibinfo {year} {2023})}\BibitemShut
  {NoStop}%
\bibitem [{\citenamefont {Cats}\ \emph {et~al.}(2021)\citenamefont {Cats},
  \citenamefont {Kuipers}, \citenamefont {De~Wind}, \citenamefont {Van~Damme},
  \citenamefont {Coli}, \citenamefont {Dijkstra},\ and\ \citenamefont
  {Van~Roij}}]{catsMachinelearningFreeenergyFunctionals2021}%
  \BibitemOpen
  \bibfield  {author} {\bibinfo {author} {\bibfnamefont {P.}~\bibnamefont
  {Cats}}, \bibinfo {author} {\bibfnamefont {S.}~\bibnamefont {Kuipers}},
  \bibinfo {author} {\bibfnamefont {S.}~\bibnamefont {De~Wind}}, \bibinfo
  {author} {\bibfnamefont {R.}~\bibnamefont {Van~Damme}}, \bibinfo {author}
  {\bibfnamefont {G.~M.}\ \bibnamefont {Coli}}, \bibinfo {author}
  {\bibfnamefont {M.}~\bibnamefont {Dijkstra}},\ and\ \bibinfo {author}
  {\bibfnamefont {R.}~\bibnamefont {Van~Roij}},\ }\bibfield  {title} {\bibinfo
  {title} {Machine-learning free-energy functionals using density profiles from
  simulations},\ }\bibfield  {journal} {\bibinfo  {journal} {APL Mater.}\
  }\textbf {\bibinfo {volume} {9}},\ \href {https://doi.org/10.1063/5.0042558}
  {10.1063/5.0042558} (\bibinfo {year} {2021}),\ \Eprint
  {https://arxiv.org/abs/2101.01942} {arxiv:2101.01942} \BibitemShut {NoStop}%
\bibitem [{\citenamefont {Samm{\"u}ller}\ and\ \citenamefont
  {Schmidt}(2024)}]{sammullerNeuralDensityFunctionals2024}%
  \BibitemOpen
  \bibfield  {author} {\bibinfo {author} {\bibfnamefont {F.}~\bibnamefont
  {Samm{\"u}ller}}\ and\ \bibinfo {author} {\bibfnamefont {M.}~\bibnamefont
  {Schmidt}},\ }\bibfield  {title} {\bibinfo {title} {Neural density
  functionals: {{Local}} learning and pair-correlation matching},\ }\href
  {https://doi.org/10.1103/PhysRevE.110.L032601} {\bibfield  {journal}
  {\bibinfo  {journal} {Phys.\ Rev.\ E}\ }\textbf {\bibinfo {volume} {110}},\
  \bibinfo {pages} {L032601} (\bibinfo {year} {2024})}\BibitemShut {NoStop}%
\bibitem [{\citenamefont {Hansen}\ and\ \citenamefont
  {McDonald}(2013)}]{hansenTheorySimpleLiquids2013}%
  \BibitemOpen
  \bibfield  {author} {\bibinfo {author} {\bibfnamefont {J.-P.}\ \bibnamefont
  {Hansen}}\ and\ \bibinfo {author} {\bibfnamefont {I.~R.}\ \bibnamefont
  {McDonald}},\ }\bibfield  {title} {\bibinfo {title} {Theory of {{Simple
  Liquids}}},\ }in\ \href {https://doi.org/10.1016/B978-0-12-387032-2.00001-5}
  {\emph {\bibinfo {booktitle} {Theory of {{Simple Liquids}}}}}\ (\bibinfo
  {publisher} {Elsevier},\ \bibinfo {year} {2013})\BibitemShut {NoStop}%
\end{thebibliography}%

\clearpage
\onecolumngrid

\begin{center}
    {\large \bfseries End Matter: Details of Pair-Correlation Matching}
\end{center}

\twocolumngrid 

In this work, we train a convolutional neural network to directly approximate the excess free energy $\mathscr{F}_{\text{exc}}[\rho]$, enabling the straightforward calculation of functional derivatives by (auto-)differentiating the neural functional with respect to its inputs. Rather than approximating the free energy directly, we train the neural functional by optimizing an objective that matches the Hessian of the network to a direct correlation function that is computed from short
simulations of homogeneous bulk systems. We refer to this approach as pair-correlation matching. 

We focus on 3D systems in a planar geometry, where the excess free energy of a system of area $A$ is a functional of the density $\rho(z)$, which is constant across any plane parallel to the $xy$-plane, i.e. $\rho(z) = \rho(x,y,z)$ with $\rho(x, y, z) = \rho(x', y', z)$ for all $(x, y), (x', y')$ within the confines of $A$. We represent the excess free-energy functional $\mathscr{F}_{\text{exc}}[\rho]$ as a neural network $F_\theta^{(2)}(\rho_1,\cdots\rho_n)$, with $\rho_i$ the density at grid point $z_i$ for $i \in \{1, \ldots, n\}$, network parameters $\theta$, and where the upper index `$(2)$' indicates that the neural network is optimized for pair correlations of bulk systems as obtained from simulations.

The functional derivative $\delta \mathscr{F}_{\text{exc}}/\delta \rho$ evaluated at $z_i$ is defined as the limit of the partial derivative  $\lim _{\Delta z \rightarrow 0} (1/\Delta z) \partial \mathscr{F}_{\text{exc}} / \partial \rho_i$ with $\rho_i = \rho(z_i)$ and $\Delta z$ the (uniform) grid spacing. We leverage this relationship by employing automatic differentiation (autodiff) to approximate the first and second functional derivatives of $\mathscr{F}_{\text{exc}}[\rho]$ on a finite grid by $(1/\Delta z) \partial F_{\theta}^{(2)}/\partial \rho_i$ and $(1/\Delta z)^2 \partial^2 F_{\theta}^{(2)}/\partial \rho_i \partial \rho_j$, respectively. We target the approximation of the second functional derivative in homogeneous bulk systems with bulk densities $\rho_b$ as our primary optimization objective, i.e. we minimize the difference between $(1/\Delta z)^2 \partial^2 F_{\theta}^{(2)}/\partial \rho_i \partial \rho_j|_{\rho_b}$ and $\delta^2 \mathscr{F}_{\text{exc}}/\delta \rho(z_i) \delta \rho(z_j)|_{\rho_b}$. 

To train our neural functional according to this optimization objective, we require ground-truth examples of $\delta^2 \mathscr{F}_{\text{exc}}/\delta \rho(z_i) \delta \rho(z_j)|_{\rho_b}$. We obtain these examples by sampling the radial distribution function $g(r)$ from simulations of homogeneous bulk systems and applying the Ornstein-Zernike equation to obtain $\delta^2 \mathscr{F}_{\text{exc}}/\delta \rho(z_i) \delta \rho(z_j)|_{\rho_b}$ from $g(r)$. Key to this transformation is the two-body direct correlation function, which is defined in terms of the second functional derivative as
\begin{equation}
    \label{eq:directcorr}
    c^{(2)}(\mathbf{r}, \mathbf{r}')
    =
    - \beta 
    \frac{\delta^2 \mathscr{F}_{\text{exc}}[\rho]}
         {\delta \rho(\mathbf{r}) \delta \rho(\mathbf{r}')},
\end{equation}
for systems with arbitrary geometry. In a uniform and isotropic bulk fluid, $c^{(2)}(\mathbf{r}, \mathbf{r}') = c_b^{(2)}(r)$ only depends on the distance between two points, 
\begin{equation}
    r = |\mathbf{r}-\mathbf{r}'| = \sqrt{(x-x')^2+(y-y')^2+(z-z')^2}.
\label{eq:r}   
\end{equation}

In such a system, we obtain the direct correlation function $c_b^{(2)}(r)$ from the radial distribution function $g(r)$ by first calculating the total correlation function $h(r)=g(r)-1$ and then applying  the bulk Ornstein-Zernike equation
\begin{equation}
    c^{(2)}_b(r) = \frac{1}{2\pi^2}\int_0^\infty \frac{\sin(kr)}{kr}\left(\frac{\hat{h}(k)}{1 + \rho \hat{h}(k)}\right)k^2dk,
\label{eq:g_to_c2}
\end{equation}
with $\hat{h}(k)$ the Fourier transform of the total correlation function $h(r)$. Since we consider systems in planar geometry, we integrate Eq.~(\ref{eq:directcorr}) with respect to the $x$ and $y$ coordinates and define polar coordinates $R=\sqrt{(x-x')^2+(y-y')^2}$ to obtain an expression for the laterally integrated direct correlation function $\bar{c}_b^{(2)}(z)$ in terms of $c_b^{(2)}(r)$, i.e.,
\begin{align}
    \begin{split}
    &~~\bar{c}_b^{(2)}(|z-z'|) = \frac{1}{A}\int dx \, dy \int dx' dy' \: c_b^{(2)}(|\mathbf{r} - \mathbf{r}'|) 
    \\
    &~~= \int_0^{\infty} dR \: 2\pi R \: c_b^{(2)}(\sqrt{R^2 + (z-z')^2})
    \\
    &~~= \int_{|z-z'|}^{\infty} dr \: 2\pi r \: c_b^{(2)}(r)
    \\
    &~~=  \frac{-\beta}{A} \frac{\delta^2 \mathscr{F}_{\text{exc}}[\rho]}{\delta \rho(z) \delta \rho(z')}.
    \end{split}
\label{eq:c2z}
\end{align}
Here we assume that $g(r)$ has converged to unity and that $c_b^{(2)}(r)$ is sufficiently short-ranged such that it has essentially decayed to zero at distance of half the box size, $r=L/2$. For the cubic systems with edge length $L=10\sigma$, we found this condition to hold for bulk densities $\rho_b\sigma^3<0.67$.

With Eq.~(\ref{eq:c2z}), we have arrived at an expression that enables us to construct a training set, by computing $\delta^2 \mathscr{F}_{\text{exc}}/\delta \rho(z_i) \delta \rho(z_j)|_{\rho_b}$ from sampled $g(r)$ through $c_b^{(2)}(r)$:
\begin{equation}
    \frac{-\beta}{A}\frac{\delta^2 \mathscr{F}_{\text{exc}}[\rho]}{\delta \rho(z_i) \delta \rho(z_j)} = \int_{|z_i-z_j|}^{\infty} \hspace{-6mm}dr \: 2\pi r \: c_b^{(2)}(r).
\label{eq:c2_2_partial}
\end{equation}

We now construct an optimization objective that minimizes the distance between $(1/\Delta z)^2\partial^2 F^{(2)}_{\theta}/\partial \rho_i \partial \rho_j$ and $\delta^2 \mathscr{F}_{\text{exc}}/\delta \rho(z_i) \delta \rho(z_j)$, i.e.,
\begin{eqnarray}
    L(\theta) &=& \nonumber \\
    & &\hspace{-13mm} \frac{1}{nm} \sum_{i,j}^{n,m} \left(\bar{c}_b^{(2)}(|z_i-z_j|) + \frac{\beta}{A(\Delta z)^2}\frac{\partial^2 F_\theta^{(2)}}{\partial \rho_i \partial \rho_j}\right)^2,
\end{eqnarray}
where $n$ denotes the number of grid points and $m$ denotes the number of Hessian rows fitted per loss evaluation. To reduce computational cost during training, we compute a uniformly sampled batch of $m=10$ rows of the Hessian $\{\{\delta^2 F^{(2)}_{\theta}/\delta \rho_i \delta \rho_j\}_{i=1}^n\}_{j=1}^m$  per loss evaluation. Optimizing for $\delta^2 \mathscr{F}_{\text{exc}}/\delta \rho(z_i) \delta \rho(z_j)$ means we lack information on the integration constant $\delta\mathscr{F}_{\text{exc}}/\delta \rho_0 = C$, akin to missing the integration constant $C$ when integrating $df/dx$ to find $f(x)$, e.g.,
\begin{equation}
    f(x) = \int dx \frac{df}{dx} + C. 
\end{equation}

Therefore, we also apply an additional offset loss term during training that minimizes the difference between the uniform $\delta \mathscr{F}_{\text{exc}}/\delta \rho(z_i)|_{\rho_b}$ corresponding to bulk densities $\rho_b$ and $\partial F^{(2)}_{\theta}/\partial \rho_i|_{\rho_b}$, given as
\begin{equation}
    L_{\text{offset}}= \frac{1}{n}\sum_{i} \left( \frac{1}{\Delta z} \cdot \frac{\partial F^{(2)}_{\theta}}{\partial \rho_i} - \frac{\delta \mathscr{F}_{\text{exc}}}{ \delta \rho(z_i)}\Big|_{\rho_b} \right)^2,
\end{equation}
where $n$ denotes the number of grid points and $\delta \mathscr{F}_{\text{exc}} / \delta \rho(z_i)$ is obtained from sampled bulk densities according to Eq.~(\ref{eq:equil_dens}). This implies that the neural functional $F_\theta^{(2)}$ actually also learns the bulk density $\rho_b$ (or actually its deviation from the ideal-gas density) at given $\mu$ and is thus expected to be able to accurately represent the bulk relation $\mu(\rho_b)$.

It is important to mention that we similarly cannot obtain the integration constant $\mathscr{F}_{\text{exc}}[0]=C$ from $\delta\mathscr{F}_{\text{exc}}/\delta \rho(z_i)$ in the case we seek access to $\mathscr{F}_{\text{exc}}[\rho]$ after training. However, since we know the low-density limit $\mathscr{F}_{\text{exc}}[0]=0$, we can correct the output of our neural network functional such that $F^{(2)}_{\theta}[\rho] - C$  approximates  $\mathscr{F}_{\text{exc}}[\rho]$, with $C = F^{(2)}_{\theta}[0]$. 

The training procedure of pair-correlation matching is summarized in Algorithm \ref{al:paircorr}. Here, we represent $c_b^{(2)}(z_i)$ numerically as $c^{(2)}_{b,i}$ with $c^{(2)}_{b,i}=c^{(2)}_{b,n-i}$ due to periodic boundary conditions. Similarly, we represent the sampled bulk density $\rho_b(z_i)$ numerically as $\rho_{b,i}$, where $\rho_{b,i} = \rho_{b,j} \text{, } \forall \, i,j \in n$. 

We construct a dataset from simulations of $10^9$ trial moves in a cubic box with an edge length of $10\sigma$ subject to periodic boundary conditions and a $\sigma/32$ grid-spacing. All simulations are carried out at distinct chemical potentials $\beta\mu \in [-4, 0.5]$, resulting in a maximum bulk density of $\rho_b\sigma^3 = 0.67$ and a minimum bulk density of $\rho_b\sigma^3 = 0.02$ within the training set. Here, we verify that $c_b^{(2)}(r)$ has essentially decayed to zero at $r=L/2$ for bulk densities $\rho_b\sigma^3<0.67$.

We employ a convolutional neural network with periodic and dilated convolutions, each  with a kernel size of 3, a dilation of 2, and 6 layers. The number of channels per layer is set to $N_{\text{channels}} =[16,16,32,32,64,64]$, applying average-pooling with kernel size 2 after each layer. This network takes as input an array of $n=320$ values of $\{\rho_i\}_{i=1}^{n}$ per datapoint and produces a single scalar output, $F^{(2)}_{\theta}$. The model is trained for 180 epochs, taking approximately   $\sim 30$ minutes on an  Nvidia RTX 4070 GPU. \\

\begin{algorithm}[H]
\caption{Pair-Correlation Matching}
\label{al:al_c2}
\KwData{$\mathcal{D} = \{\{c^{(2),0}_{b,i}, \rho^{0}_{b,i}\}_{i=1}^n, \ldots, \{c_{b,i}^{(2),D}, \rho^{D}_{b,i}\}_{i=1}^n\}$ containing $D$ pair-correlation functions $\{c^{(2)}_{b,i} \}_{i=1}^{n}$; $D$ bulk density profiles $\{\rho_{b,i}\}_{i=1}^n$; loss scaling factor $\alpha=1/1000$; loss scaling factor $\beta=1/32$; number of uniformly sampled Hessian rows $m=10$.} 
\KwResult{trained neural functional $F^{(2)}_{\theta}(\{\rho_{i}\}_{i=1}^n)$. }
\For{epoch}{
    \For{each $\{c^{(2)}_{b,i}, \rho_{b,i}\}_{i=1}^n$ in $\mathcal{D}$}{
      compute NN output $F^{(2)}_{\theta}(\{\rho_{b,i}\}_{i=1}^n)$\;
      compute $\{\partial F^{(2)}_{\theta} / \partial \rho_{i}\}_{i=1}^n$ with autodiff\;
      uniformly sample batch $\mathcal{B} = \{\partial F^{(2)}_{\theta} / \partial \rho_i\}_{i=1}^{m}$ from $\{\partial F^{(2)}_{\theta} / \partial \rho_{i}\}_{i=1}^n$\;
      \For{each $\partial F^{(2)}_{\theta} / \partial \rho_i$ in $\mathcal{B}$}{
        compute $\{\partial^2 F^{(2)}_{\theta} / \partial \rho_i \partial \rho_j\}_{j=1}^n$ with autodiff\;
        compute $\{\delta^2 \mathscr{F}_{\text{exc}}/ \delta \rho(z_i) \delta \rho(z_j) \}_{j=1}^n$ from $\{c^{(2)}_{b,i} \}_{i=1}^{n}$ with Eq.~(\ref{eq:c2_2_partial})\;
      }
      $L(\theta) = \frac{1}{nm}\sum_{i=1}^m\sum_{j=1}^n (\delta^2 \mathscr{F}_{\text{exc}}/ \delta \rho(z_i) \delta \rho(z_j) - (1/\Delta z)^2 \partial^2 F^{(2)}_{\theta} / \partial \rho_i \partial \rho_j)^2$\;
      compute $\{\delta \mathscr{F}_{\text{exc}}/ \delta \rho(z_i) \}_{i=1}^n$ with Eq.~(\ref{eq:equil_dens})\;
      $L_{\text{offset}}(\theta) = \frac{1}{n}\sum_{i=1}^n \left(\delta \mathscr{F}_{\text{exc}}/ \delta \rho(z_i) - (1/\Delta z) \partial F^{(2)}_{\theta} / \partial \rho_i \right)^2$\;
      $L(\theta) = \alpha \cdot L(\theta) + \beta \cdot L_{\text{offset}}(\theta)$\;  
      update parameters $\theta \leftarrow \text{Optimizer}(\theta, \nabla_\theta L(\theta)$\;
    }
}
\label{al:paircorr}
\end{algorithm}

\end{document}


\title{Supplementary Material: Learning Neural Free-Energy Functionals with Pair-Correlation Matching}

\date{\today}

\author{Jacobus Dijkman}
\affiliation{Van 't Hoff Institute for Molecular Sciences, University of Amsterdam, The Netherlands}
\affiliation{Informatics Institute, University of Amsterdam, The Netherlands}   
\author{Marjolein Dijkstra}
\affiliation{Soft Condensed Matter \& Biophysics, Debye Institute for Nanomaterials Science, Utrecht University, The Netherlands}
\author{Ren\'{e} van Roij}
\affiliation{Institute for Theoretical Physics, Utrecht University, The Netherlands}
\author{\\Max Welling}
\affiliation{Informatics Institute, University of Amsterdam, The Netherlands}
\author{Jan-Willem van de Meent}
\affiliation{Informatics Institute, University of Amsterdam, The Netherlands}
\author{Bernd Ensing}
\affiliation{Van 't Hoff Institute for Molecular Sciences, University of Amsterdam, The Netherlands}
\affiliation{AI4Science Laboratory, University of Amsterdam, The Netherlands}

\maketitle

\onecolumngrid 
\vspace{-2em}
\tableofcontents
\vspace{2em}
\twocolumngrid 


\section{1. Classical Density Functional Theory}

Classical density functional theory (cDFT) is a grand-canonical framework that relies on the fact that the variational grand potential $\Omega[\rho]$ of a classical one-component many-body system at a given temperature $T$ is uniquely determined by the particle density $\rho(\mathbf{r})$ via 
\begin{equation}
\Omega[\rho]= \mathscr{F}[\rho] + \int \mathrm{d} r \rho(\mathbf{r})\left(V_{\text {ext }}(\mathbf{r})-\mu\right),
\label{eq:grand_pot}
\end{equation}
with $\mathscr{F}[\rho]$ representing the intrinsic Helmholtz free-energy functional, $V_{\text{ext}}(\mathbf{r})$ the external potential, and $\mu$ the chemical potential. For a given particle-particle interaction and temperature, the unique density functional $\mathscr{F}[\rho]$ determines the thermodynamic and structural equilibrium properties of a system for any chemical potential and external potential. Within cDFT, it is conventional  to split the intrinsic free-energy functional into an ideal and  excess contribution, $\mathscr{F}[\rho]= 
\mathscr{F}_{\text{id}}[\rho] + \mathscr{F}_{\text{exc}}[\rho]$. The ideal-gas contribution is exactly known as 
\begin{equation}
    \mathscr{F}_\text{id}[\rho] = \frac{1}{\beta} \int \operatorname{dr} \rho(\mathbf{r})\big(\ln \rho(\mathbf{r})\Lambda^3-1\big),
\label{eq:eq_dens}
\end{equation}
with $\beta=1/k_BT$ and $\Lambda$  the thermal wavelength. 

Mathematical proofs exist \cite{evansNatureLiquidvapourInterface1979} stating that (i) the equilibrium density profile, denoted here as  $\rho_0({\bf r})$,  minimizes $\Omega[\rho]$, and (ii) the equilibrium grand potential equals $\Omega[\rho_0]$. Clearly, once $\mathscr{F}[\rho]$  for the system of interest is known, the Euler-Lagrange equation
$\delta\Omega[\rho]/\delta\rho({\bf r})|_{\rho_0}=0$ can be solved to find $\rho_0({\bf r})$ and $\Omega[\rho_0]$. The Euler-Lagrange equation takes the form
\begin{equation}
    \rho_0(\mathbf{r})= \frac{1}{\Lambda^{3}}\exp \left(\beta\mu-\left.\beta \frac{\delta \mathscr{F}_{\text {exc}}[\rho]}{\delta \rho(\mathbf{r})}\right|_{\rho=\rho_0}\hspace{-5mm}-\beta V_{\text {ext }}(\mathbf{r})\right).
\label{eq:equil_dens}
\end{equation}
This self-consistency relation can be leveraged to find $\rho_0(\mathbf{r})$ through Picard iteration \cite{rothIntroductionDensityFunctional2006, edelmannNumericalEfficientWay2016, mairhoferNumericalAspectsClassical2017}.


\section{2. Training on the One-Body Direct Correlation Function \label{section:c1learning}}

\begin{figure*}
\includegraphics[width=0.8\textwidth]{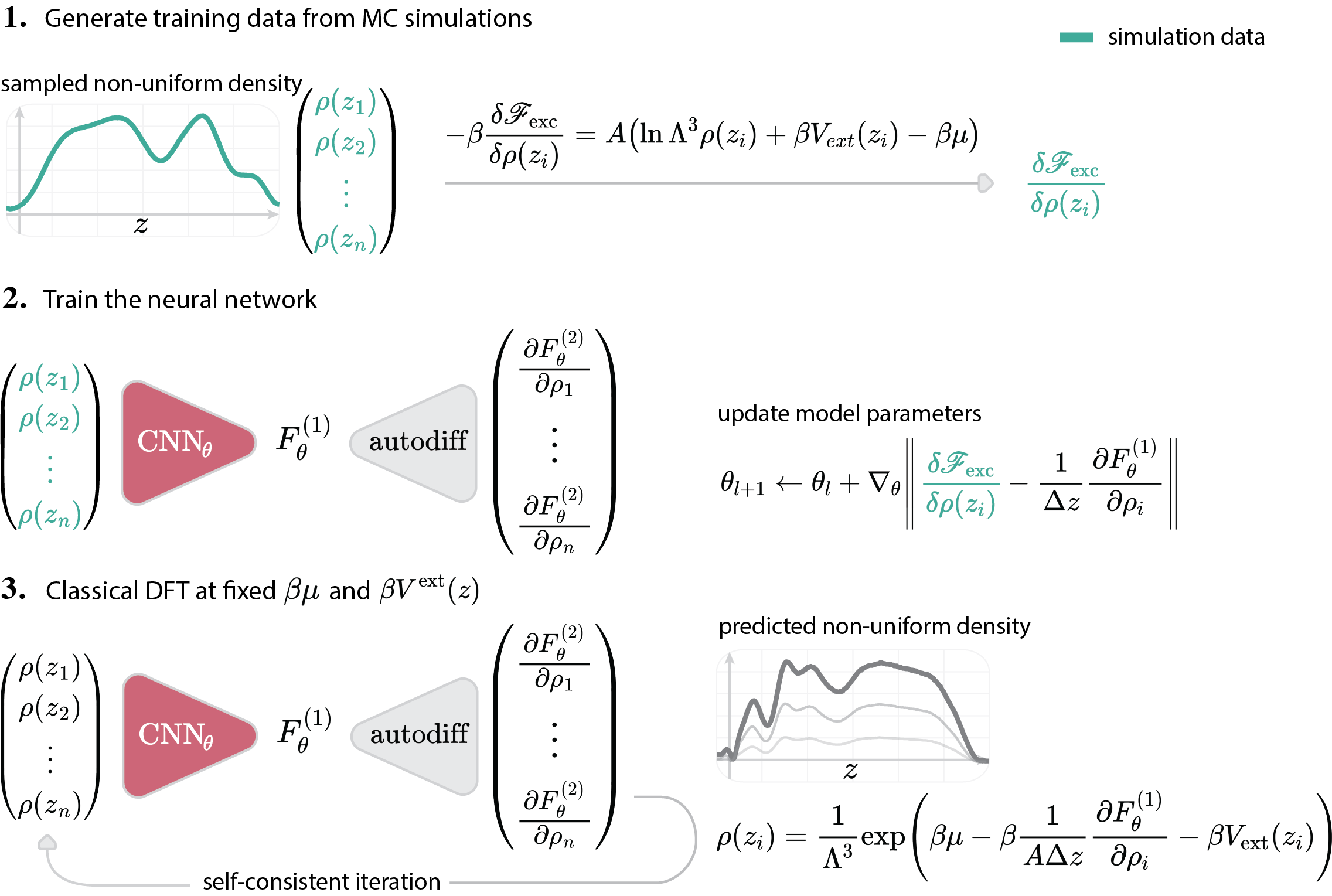}
\caption{\small The neural free-energy functional $F_\theta^{(1)}$ is trained by fitting the gradient of the model output with respect to the input to the first functional derivative of the excess free energy as obtained from simulation, after which it can be applied within the classical Density Functional Theory (cDFT) framework. \textbf{1.} Non-uniform densities are sampled from Monte Carlo simulations of Lennard-Jones particles subjected to inhomogeneous external potentials, from which $\delta \mathscr{F_{\text{exc}}}/ \delta \rho(z_i)$ is derived. \textbf{2.} Through automatic differentiation (autodiff), the neural functional is optimized to fit the gradient of the model output with respect to input density profiles to $\delta \mathscr{F_{\text{exc}}}/ \delta \rho(z_i)$. \textbf{3.} The optimized model can then be applied in cDFT to obtain the non-uniform equilibrium densities and the excess free energy for a system of Lennard-Jones particles subjected to inhomogeneous external potentials.}
\label{fig:c1_training}
\end{figure*}

Instead of using pair-correlation matching, we can train a neural free-energy functional $F_{\theta}^{(1)}$ by minimizing the error between $\delta \mathscr{F}_{\text{exc}}/\delta \rho(z_i)$ and $(1/\Delta z) \partial F_{\theta}^{(1)}/\partial \rho_i$, as illustrated in Fig.~\ref{fig:c1_training}. We denote this neural functional as $F_{\theta}^{(1)}$, as the first functional derivative of the excess free energy is associated with the one-body direct correlation function by $c^{(1)}(x,y,z) = - \beta \delta \mathscr{F}_{\text{exc}}/\delta \rho(x,y,z)$. Relating to previous work, this neural functional is different from the neural functional of the one-body direct correlation function developed by \textcite{sammullerNeuralFunctionalTheory2023a}, as expanded upon in Section 8 of the Supplementary Material.

We obtain the first functional derivative of the excess free energy from the equilibrium density by rearranging Eq.~(\ref{eq:equil_dens}):
\begin{eqnarray}
- \beta \frac{\delta \mathscr{F}_{\text{exc}}}{\delta \rho(x,y,z)}& =& \nonumber \\
& & \hspace{-15mm}\ln \Lambda^3 \rho (x,y,z) + \beta V_{ext}(x,y,z) - \beta \mu.
\label{eq:dFdrho}
\end{eqnarray}

We consider 3D systems in a planar geometry, where the excess free energy of a system with area $A$ is a functional of the density $\rho(z)$, which is constant across any plane parallel to the $xy$-plane, i.e., $\rho(z) = \rho(x,y,z)$ with $\rho(x, y, z) = \rho(x', y', z)$ for all $(x, y), (x', y')$ within the confines of $A$. For such a system, we can write
\begin{equation}
- \beta \frac{\delta \mathscr{F}_{\text{exc}}}{\delta \rho(z)} = A(\ln \Lambda^3 \rho (z) + \beta V_{ext}(z) - \beta \mu).
\label{eq:dFdrho}
\end{equation}
This means that we can obtain $\delta \mathscr{F}_{\text{exc}}/\delta \rho(z)$ by sampling equilibrium densities from simulation. We sample density profiles of inhomogeneous systems of Lennard-Jones particles above the critical point at a temperature $k_BT/\epsilon = 2$. We construct a dataset from simulations of $10^9$ trial moves in a cubic box with an edge length of $10\sigma$ subject to periodic boundary conditions and a $\sigma/32$ grid-spacing. All simulations are conducted at  distinct chemical potentials $\beta\mu \in [-4, 0.5]$ and a maximum local density of $\rho(z)\sigma^3 = 0.67$. The choice of external potentials is explained in Section 3 of the Supplementary Material. We use the same convolutional neural network architecture  as for the neural functional $F_\theta^{(2)}$: a convolutional neural network with periodic and dilated convolutions, each  with a kernel size of 3, a dilation of 2, and 6 layers. The number of channels per layer is set to $N_{\text{channels}} =[16,16,32,32,64,64]$, applying average-pooling with kernel size 2 after each layer. The model is trained for 5000 epochs, requiring approximately  $100$ minutes on an Nvidia RTX 4070 GPU. The training procedure is summarized in Algorithm \ref{al:al_c1}. 

Since this method is dependent on the set of inhomogeneous densities included in the train dataset, the accuracy of the $F_\theta^{(1)}$ functional could plausibly be improved further by iterating on the design of the training dataset. We have not tested this extensively, since our primary interest in this work is in the pair correlation matching methodology.

\begin{algorithm}[h]
\caption{Training on $\delta \mathscr{F}_{\text{exc}}/ \delta \rho (z)$}
\label{al:al_c1}
\KwData{train dataset $\mathcal{D} = \{\{\rho^0_i\}_{i=1}^n, \ldots, \{\rho^D_i\}_{i=1}^n\}$ consisting of $D$ density profiles $\{\rho_i\}_{i=1}^n$ evaluated at gridpoints $\{z_i \}_{i=1}^n$.} 
\KwResult{trained neural network model $F^{\text{exc}}_\theta(\{\rho_i\}_{i=1}^n)$.}
\For{epoch}{
    \For{each $\{\rho_i\}_{i=1}^n$ in $\mathcal{D}$}{
      generate model output scalar $F^{(1)}_\theta(\{\rho_i\}_{i=1}^n)$\;
      compute $\{\partial F^{(1)}_\theta / \partial \rho_i\}_{i=1}^n$ with autodiff\;
      compute $\{\delta \mathscr{F}_{\text{exc}}/ \delta \rho(z_i) \}_{i=1}^n$ with Eq.~(\ref{eq:dFdrho})\;
      $L_\theta = \frac{1}{n}\sum_{i=1}^n (\delta \mathscr{F}_{\text{exc}}/ \delta \rho(z_i) - (1/\Delta z) \partial F^{(1)}_\theta / \partial \rho_i)^2$\;
      update parameters $\theta \leftarrow \text{Optimizer}(\theta, \nabla_\theta L_\theta)$\;
    }
}
\label{al:c1_training}
\end{algorithm}


\section{3. Construction of External Potentials}
The inhomogeneous density profiles used in this work, both for  performance evaluation and  training of the neural functional $F_\theta^{(1)}$, were generated via MC simulations subjected to various external potentials. These external potentials consist of randomized variations of well-potentials and Gaussian potentials. 

The form of the well-potentials is adapted from \textcite{catsMachinelearningFreeenergyFunctionals2021}, where  
\begin{equation}
\beta V_{\text {well}}(z)=\left\{\begin{array}{lll}
0 & \text { for } & |z| \in [-w\frac{L}{2},w\frac{L}{2}], \\
s\left(\frac{|z|-w \frac{L}{2}}{(1-w) \frac{L}{2}}\right)^p & \text { for } & |z|>w \frac{L}{2}.
\end{array}\right.
\label{eq:well}
\end{equation}
Here,  $s$ represents the dimensionless strength characterizing  the potential at $|z| = L/2$, uniformly sampled with $s \sim \mathcal{U}(40, 60)$;  $w$ denotes the width of the central part of the slit ($\beta V_{\text{ext}}=0$) and was uniformly sampled with $w \sim \mathcal{U}(0.4, 0.9)$;  $p$ characterizes the steepness of the potential and was uniformly sampled with $p \sim \mathcal{U}(2, 9)$. 

The Gaussian potentials were constructed as a sum of Gaussians, expressed as 
\begin{equation}
\beta V_{\text{Gauss}}(z)=\sum_{i=1}^{N} h_i \exp \left(-\frac{\left(z-\mu_i\right)^2}{2 \sigma_i^2}\right),
\label{eq:gauss}
\end{equation}
where the number of Gaussians $N$ is randomly chosen between $N=0$ and $N=10$; the mean of the Gaussians $\mu_i$ is uniformly sampled from $\mu \sim \mathcal{U}(0, L)$; the standard deviation of the Gaussians $\sigma_i$ is uniformly sampled from $\sigma \sim \mathcal{U}(0, L/10) + L/100$; the height of the Gaussians $h_i$ is randomly sampled from a squared normal distribution, $h \sim \mathcal{U}^2(\mu=0, \sigma^2=1)$.

The potentials constructed from a combination of well-potentials and Gaussian potentials were simply constructed by summing the individual potentials 
\begin{equation}
    \beta V_{\text{ext}}(z) = \beta \left(V_{\text{well}}(z) + V_{\text{Gauss}}(z)\right).
\end{equation}

For the results shown in Fig.~2 of the main text, simulations were performed for 150 distinct external potentials, of which 50 densities were generated using pure well-potentials, 50 were generated using a set of Gaussian potentials and 50 were generated using a  combination of well-potentials and Gaussian potentials.

To train the neural functional $F_\theta^{(1)}$, a distinct dataset was constructed by using 200 pure well-potentials; 500 Gaussian potentials; 300 combinations of well-potentials and  Gaussian potentials; 100 systems without an external potential. All simulations were conducted at a randomly selected chemical potential $\beta \mu \in [-4,0.5]$.


\section{4. Obtaining the Excess Free Energy from Simulation}

The excess Helmholtz free energy of the homogeneous bulk fluid (at a fixed volume $V$ and temperature $T$) can be obtained from grand-canonical simulation via the equilibrium grand potential $\Omega_{eq}(\mu)$. The latter can be obtained by thermodynamic integration 
\begin{equation}
    \Omega_{eq}(\mu)=-\int_{-\infty}^\mu \mathrm{d} \mu^{\prime}N(\mu^{\prime}),
\label{eq:thermo_int}
\end{equation}
where $N(\mu')=V\rho_b(\mu')$ is the simulated (average) number of particles at chemical potential $\mu'$ in this homogeneous bulk system. Using the thermodynamic relation $\Omega_{eq}=\mathscr{F}-\mu N$ and  $\mathscr{F}=\mathscr{F}_{id}+\mathscr{F}_{exc}$ with $\mathscr{F}_{id}=Nk_BT(\ln(N\Lambda^3/V)-1)$, we arrive at $\mathscr{F}_{exc}=\Omega_{eq}+\mu(N) N-\mathscr{F}_{id}$. Here $\mu(N)$ is the inverse of $N(\mu)$, and we can identify $\mathscr{F}_{exc}$ with the excess free-energy functional evaluated at the homogeneous bulk, $\mathscr{F}_{exc}[\rho_b]$. 

To calculate the integral of Eq.~(\ref{eq:thermo_int}), we sample the number of particles for a system in steps of $\Delta \beta \mu = 0.2$, for a range of $\beta\mu=-4$ (corresponding to a bulk density of $\rho_b\sigma^3 = 0.02$) up to the target chemical potential. The resulting excess free energy from simulations is favorably compared with its neural-network representations $F_\theta^{(1)}$ and $F_\theta^{(2)}$ in Fig.~2 of the main text. 


\section{5. Numerical Errors in the Radial Distribution Function}

\begin{figure}
    \centering
    \includegraphics[width=0.35\textwidth]{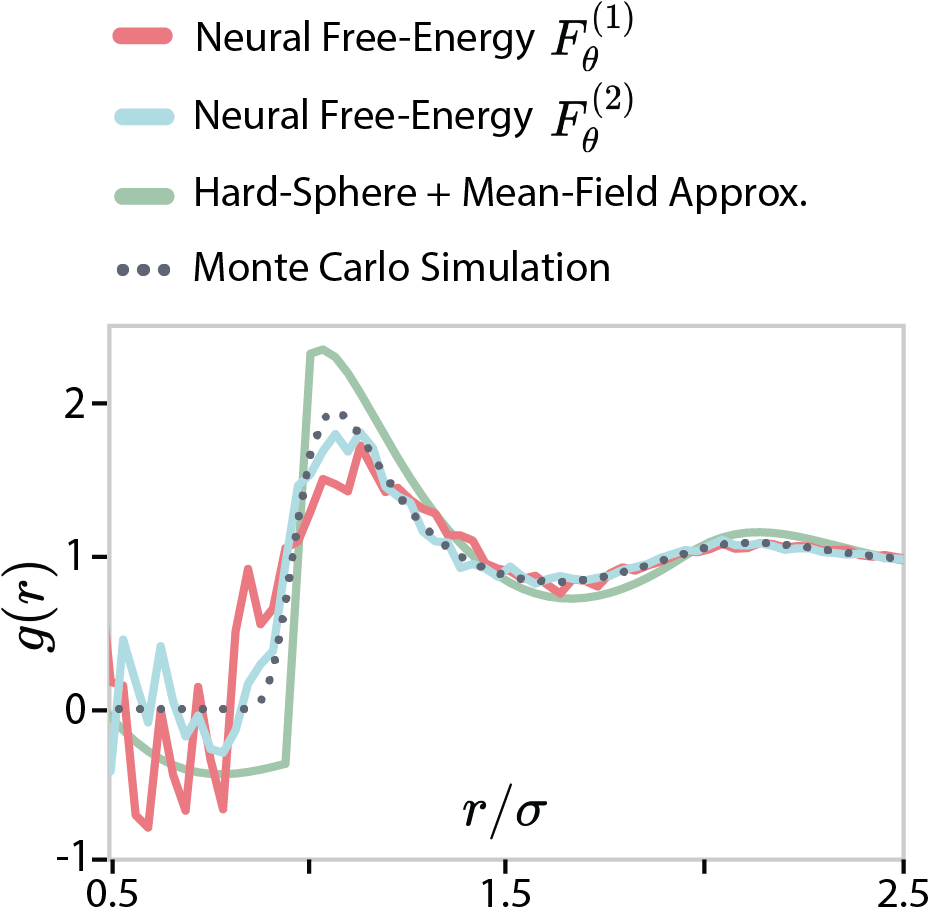}
    \caption{\small Unfiltered estimates of the radial distribution function for $\rho_b\sigma^3=0.67$. The radial distribution estimates from the neural functionals $F_\theta^{(1)}$, $F_\theta^{(2)}$ and $F^{\text{MF}}_{\text{exc}}$, which represents the analytical approximation of the White-Bear mark II version of Fundamental Measure Theory (FMT) combined with a mean-field approximation for the attractive part of the Lennard-Jones potential, are compared with simulation results.}
    \label{fig:rdf_unfiltered}
\end{figure}

The derivation of $g(r)$ from $c^{(2)}(z)$ involves several extensive transformations, which can introduce sensitivity to numerical errors. This numerical instability is most pronounced when $g(r) \rightarrow 0$, as shown in Fig.~\ref{fig:rdf_unfiltered}.

To reduce the numerical errors in this region, we firstly average the bulk $\delta^2 F_\theta/ \delta \rho(z_i) \delta \rho(z_j)$ values across rows of the neural functional Hessian, since $\delta^2 F_\theta/ \delta \rho(z_i) \delta \rho(z_j)$ should be exactly the same as $\delta^2 F_\theta/ \delta \rho(z_j) \delta \rho(z_i)$, due to the symmetry of a bulk system. In practice, small numerical differences remain between the Hessian rows for the neural functionals $F_\theta^{(1)}$, $F_\theta^{(2)}$ after training, which can influence the result of the extensive transformation from $c^{(2)}(z)$ to $g(r)$. Therefore averaging $\delta^2 F_\theta/ \delta \rho(z_i) \delta \rho(z_j)$ across the Hessian reduces numerical errors when calculating $g(r)$. In addition, we slightly adapt $g(r)$ by filtering it as%
\begin{equation}
g_{\text{filter}}(r_i) = \begin{cases} 
0 & \text{for all } r_i < r_0, \\
g(r_i) & \text{otherwise},
\end{cases}
\end{equation}
where $r_0$ is the largest interparticle distance $r_i$ for which $g(r_i) \leq 0$. 


\section{6. More Examples of Density Estimates}

\begin{figure*}
    \begin{center}
    \includegraphics[width=1\textwidth]{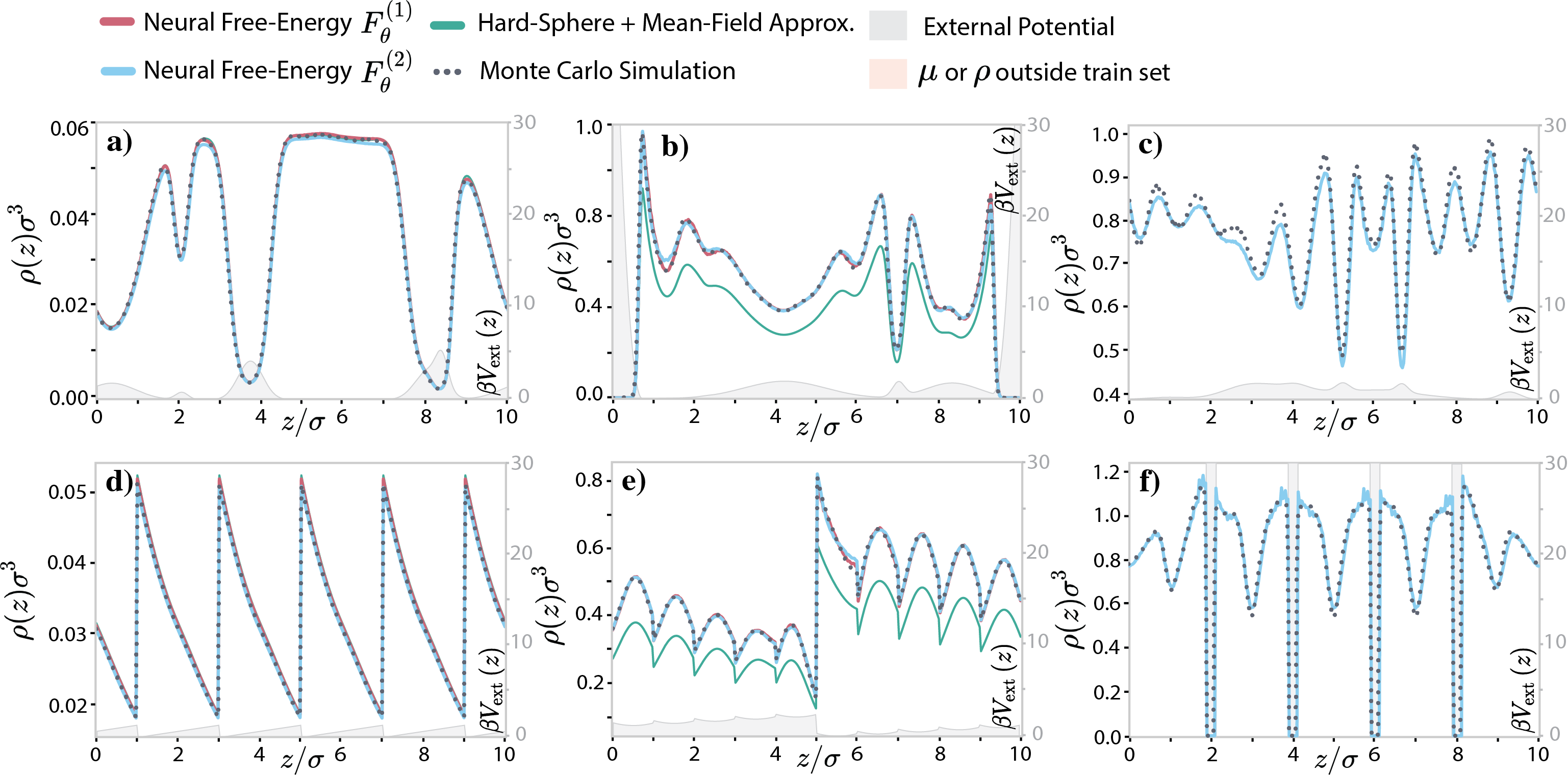}
    \end{center}
    \vspace{-1.5em} 
    \caption{\small Evaluation of the neural free-energy functionals $F_\theta^{(1)}$ and $F_\theta^{(2)}$, where $F_\theta^{(1)}$ is optimized based on one-body correlation functions  while $F_\theta^{(2)}$ is optimized using pair-correlation matching. Inhomogeneous density profiles are shown at six different external potentials (gray filling, right axes) through classical DFT comparing the functionals $F_\theta^{(1)}$, $F_\theta^{(2)}$ and $F^{\text{MF}}_{\text{exc}}$, which represents the analytical approximation of the White-Bear mark II version of Fundamental Measure Theory (FMT) combined with a mean-field approximation for the attractive part of the Lennard-Jones potential. The chemical potentials are given by \textbf{a/d)} $\beta \mu = -3$, \textbf{b/e)} $\beta \mu = 0$, and \textbf{c/f)} $\beta \mu = 3$, the latter being far beyond the training set $\beta\mu \in [-4,0.5]$ where only $F_\theta^{(2)}$ gives a converged solution.}
    \label{fig:supp_results}
\end{figure*}

In addition to Figure 2a in the main text, we present additional examples of DFT predictions for various external potentials, as shown in Fig.~\ref{fig:supp_results}. In this Figure, the columns depict increasing  chemical potentials from left to right: $\beta\mu = -3$ for Fig.~\ref{fig:supp_results}a and \ref{fig:supp_results}d; $\beta\mu = 0$ for Fig.~\ref{fig:supp_results}b and \ref{fig:supp_results}e; and $\beta\mu = 3$ for Fig.~\ref{fig:supp_results}c and \ref{fig:supp_results}f. 

The external potentials shown in Fig.~\ref{fig:supp_results}a--c are generated using a set of Gaussian potentials (Fig.~\ref{fig:supp_results}a and \ref{fig:supp_results}c), and a combination of Gaussian potentials and a well-potential (Fig.~\ref{fig:supp_results}b), as described in Section 3 of the Supplementary Material. We observe that both  neural functionals $F_\theta^{(1)}$ and $F_\theta^{(2)}$ provide accurate predictions for $\beta \mu = -3$ and $\beta\mu = 0$, both outperforming $F^{\text{MF}}_{\text{exc}}$. However, $F^{\text{MF}}_{\text{exc}}$ and $F_\theta^{(1)}$ both fail to converge to a solution for $\beta \mu = 3$ (Fig.~\ref{fig:supp_results}c and Fig.~\ref{fig:supp_results}f), which lies far outside  the training set range of $\beta \mu \in [-4, 0.5]$. While  $F_\theta^{(2)}$ is also trained with (uniform) densities up to $\beta \mu = 0.5$, it is still capable of producing relatively accurate predictions for $\beta \mu = 3$.

Additionally, we test the performance of the neural functionals on a number of atypical external potentials that are in no way related to the family of external potentials used elsewhere in  this work (Fig.~\ref{fig:supp_results}d--f). Here we observe that both  neural functionals $F_\theta^{(1)}$ and $F_\theta^{(2)}$ still perform well for these potentials for $\beta \mu = -3$ and $\beta \mu = 0$. Again,  $F^{\text{MF}}_{\text{exc}}$ and $F_\theta^{(1)}$ fail to converge to a solution for $\beta \mu = 3$, whereas  $F_\theta^{(2)}$  provides a relatively accurate estimate (Fig.~\ref{fig:supp_results}f).


\section{7. Limitations of Pair-Correlation Matching} 

\begin{figure*}
    \begin{center}
    \includegraphics[width=1\textwidth]{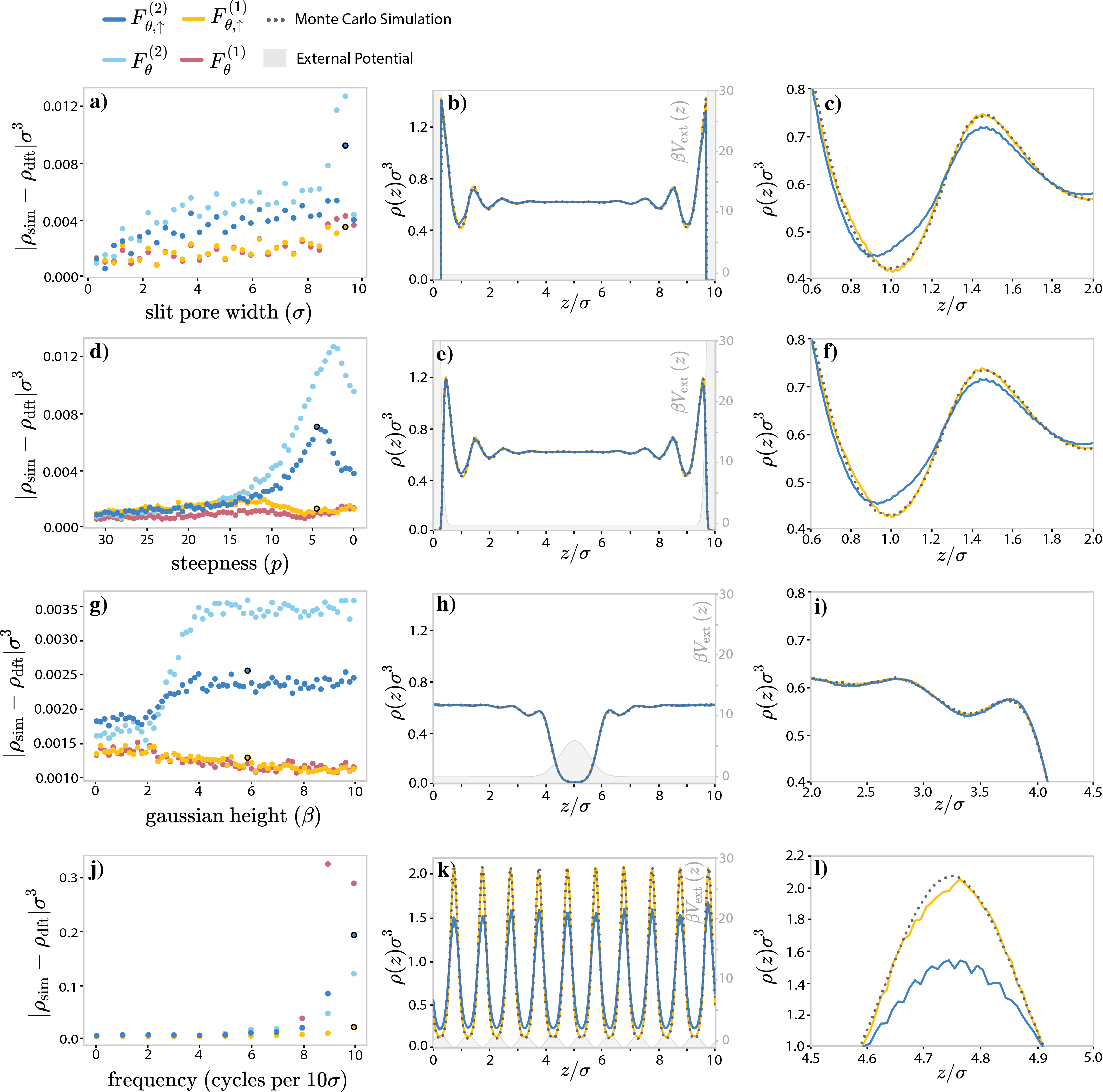}
    \end{center}
    \vspace{-1.5em} 
    \caption{\small The performance of neural functionals trained with pair-correlation matching in comparison to neural functionals trained with a dataset of inhomogeneous densities on a selection of external potentials with a large degree of inhomogeneity, specifically selected to highlight settings where neural functional predictions deviate from Monte Carlo data when increasing the inhomogeneity. All systems are subjected to $\beta \mu = 0$. Neural functional $F_{\theta}^{(1)}$ (red) is trained with inhomogeneous densities within the range $-4 < \beta \mu < 0.5$; Neural functional $F_{\theta}^{(2)}$ (light blue) is trained with bulk pair-correlation functions in the range $-4 < \beta \mu < 0.5$; Neural functional $F_{\theta, \uparrow}^{(1)}$ (yellow) is trained with inhomogeneous densities within the range $-4 < \beta \mu < 3$; Neural functional $F_{\theta, \uparrow}^{(2)}$ (dark blue) is trained with bulk pair-correlation functions in the range $-4 < \beta \mu < 3$. \textbf{a/d/g/j)} mean absolute error $\frac{1}{n} \sum\left|\rho_{\text {sim}}\left(z_i\right)-\rho_{\text{dft}}\left(z_i\right)\right|$ of DFT predictions and MC simulations of particle densities for respectively a slit-pore potential with increasing width; a wall potential with increasing steepness indicated by $p$ as specified in \ref{eq:well}; a Gaussian potential with increasing height; a sine potential with increasing frequency. Errors corresponding to the examples shown in the two rightmost columns are highlighted with a black circle. \textbf{b/e/h/k)} comparison of DFT predictions at the external potential that induced the largest mean error of the $F_{\theta,\uparrow}^{(2)}$ functional at respectively a width of $9.4\sigma$; steepness parameter $p=4.6$; Gaussian height of $5.9/\beta$; sine wavenumber equal to $2\pi/\sigma$. \textbf{c/f/i/l)} Zoom-in of the most erroneous region of the DFT estimate of the $F_{\theta,\uparrow}^{(2)}$ functional the system shown in b/e/h/k). 
    }
    \label{fig:extreme_vext}
\end{figure*}

Given the surprising ability to accurately predict inhomogeneous density profiles by learning from merely bulk systems, it is natural to ask whether there are limits to the regimes where pair correlation can be applied successfully. In particular, we would like to understand to what extent inhomogeneities can be described within this approach and how far we can extrapolate away from the bulk using the pair-correlation matching approach. Moreover, since the pair-correlation matching approach relies on the inhomogeneity of the pair-correlation function for $F_\theta^{(2)}$ to approximate inhomogeneous densities, it is reasonable to question whether including pair-correlation functions at higher bulk densities in the training set would increase the performance of the pair-correlation matching approach for predicting highly inhomogeneous densities.

To investigate the extent to which pair-correlation matching is able to accurately describe inhomogeneities, we explore 4 types of external potentials for which a clear trend is visible between systematically increasing their inhomogeneity and the impact on prediction accuracy. Figure \ref{fig:extreme_vext} illustrates these scenarios: Increasing the slit-pore width (Fig. \ref{fig:extreme_vext}a-c); Increasing the steepness of a hyperbolic tangent potential (Fig. \ref{fig:extreme_vext}d-f); Increasing the height of a Gaussian potential (Fig. \ref{fig:extreme_vext}g-i); Increasing the wave number of a sinusoidal potential (Fig. \ref{fig:extreme_vext}j-l). All systems are subjected to $\beta \mu = 0$.

We additionally investigate to what extent the predictive power of $F_\theta^{(2)}$ in highly inhomogeneous systems can be improved by  extending the bulk density range of the train set. We compare with neural functionals $F^{(1)}_{\theta,\uparrow}$, and $F^{(2)}_{\theta,\uparrow}$ which have been trained with inhomogeneous densities or bulk pair-correlation functions in the range $-4 < \beta \mu < 3$ respectively, as expanded upon in Section 9. All neural functionals investigated in this section are trained on inhomogeneous densities and pair-correlation functions with a resolution of $\Delta z = \sigma/100$, for which the details can be found in Section 10.

Fig.~\ref{fig:extreme_vext}a-c demonstrates the effect of an increasingly wide slit-pore on prediction errors. Fig. \ref{fig:extreme_vext}a shows that this increase in inhomogeneity affects the $F_{\theta}^{(2)}$-type functionals more than the $F_{\theta}^{(1)}$-type functionals. Fig. \ref{fig:extreme_vext}b shows the density profile for a barrier width of $0.625\sigma$, where $F_{\theta,\uparrow}^{(2)}$ exhibited the largest errors across the range shown in Fig. \ref{fig:extreme_vext}b. A closer examination of the most erroneous region (Fig. \ref{fig:extreme_vext}c) reveals that the $F_{\theta, \uparrow}^{(2)}$ functional underestimated the peaks and valleys of the density profile near the barrier. 

Fig.~\ref{fig:extreme_vext}d-f shows the effect of an increasingly steep wall potential on the prediction error. As seen in Fig. \ref{fig:extreme_vext}d, the error of the $F_{\theta}^{(2)}$-type functionals are affected by increasing steepness of the potential. The predicted DFT densities are shown for the external potential with the highest $F_{\theta,\uparrow}^{(2)}$ error, at $p=4.6$ in Fig.~\ref{fig:extreme_vext}e-f. As expected by the similarity in external potential, the prediction error produced in the region near the wall shown in \ref{fig:extreme_vext}f is similar to that in Fig.~\ref{fig:extreme_vext}c. 

Fig.~\ref{fig:extreme_vext}g-i shows the effect of an increasingly high Gaussian peak potential on the prediction error of the neural functionals. Since this potential is perhaps less extreme than the other potentials, the error across the domain is also lower. However, we do still observe a small discrepancy between the increase of the Gaussian height between the $F_{\theta}^{(1)}$-type and $F_{\theta}^{(1)}$-type functionals. The DFT predictions at the Gaussian potential with the highest $F_{\theta, \uparrow}^{(2)}$ prediction error (with a Gaussian height of $5.9/\beta$) are shown in Fig.~\ref{fig:extreme_vext}h-i.

Fig.~\ref{fig:extreme_vext}j-l shows the effect of increasing the wave number of a sinusoidal potential on the prediction error of the neural functionals. This scenario proved to be the most challenging for all functionals. Focusing on the $F_{\theta}^{(2)}$-type functionals, we see that the prediction error increases significantly at higher wave numbers. Fig. \ref{fig:extreme_vext}k-l shows the DFT predictions at a sinusoidal potential with a wave number  equal to $2\pi/\sigma$, which induced the highest $F_{\theta,\uparrow}^{(2)}$ prediction error. In Fig \ref{fig:extreme_vext}j, we can also see that the $F_{\theta}^{(1)}$ functional shows the highest observed prediction error across all functionals for periods getting as small as $\sigma$, whereas larger periods (so smaller wave numbers) yield better results for all functionals. This can be attributed by the fact that the $F_{\theta}^{(1)}$ functional has probably not seen these types of density fluctuations in its train set of $-4 < \beta \mu < 0.5$. Indeed, including inhomogeneous densities up to $\beta \mu = 3$ incorporates state points with higher density variations in the train set of $F_{\theta,\uparrow}^{(1)}$ and lowers the prediction error.

These results suggest that, while pair-correlation matching is generally robust, we do observe a degradation of accuracy when predicting densities for highly inhomogeneous external potentials. As might be expected, increasing the range of bulk densities used during training significantly increases the predictive performance in highly inhomogeneous external potentials. These observations also suggest that adding bulk data for $\beta \mu > 3$ could improve the performance further yet.


\section{8. Local Learning and Pair-Correlation Matching}

\begin{figure*}
\includegraphics[width=0.85\textwidth]{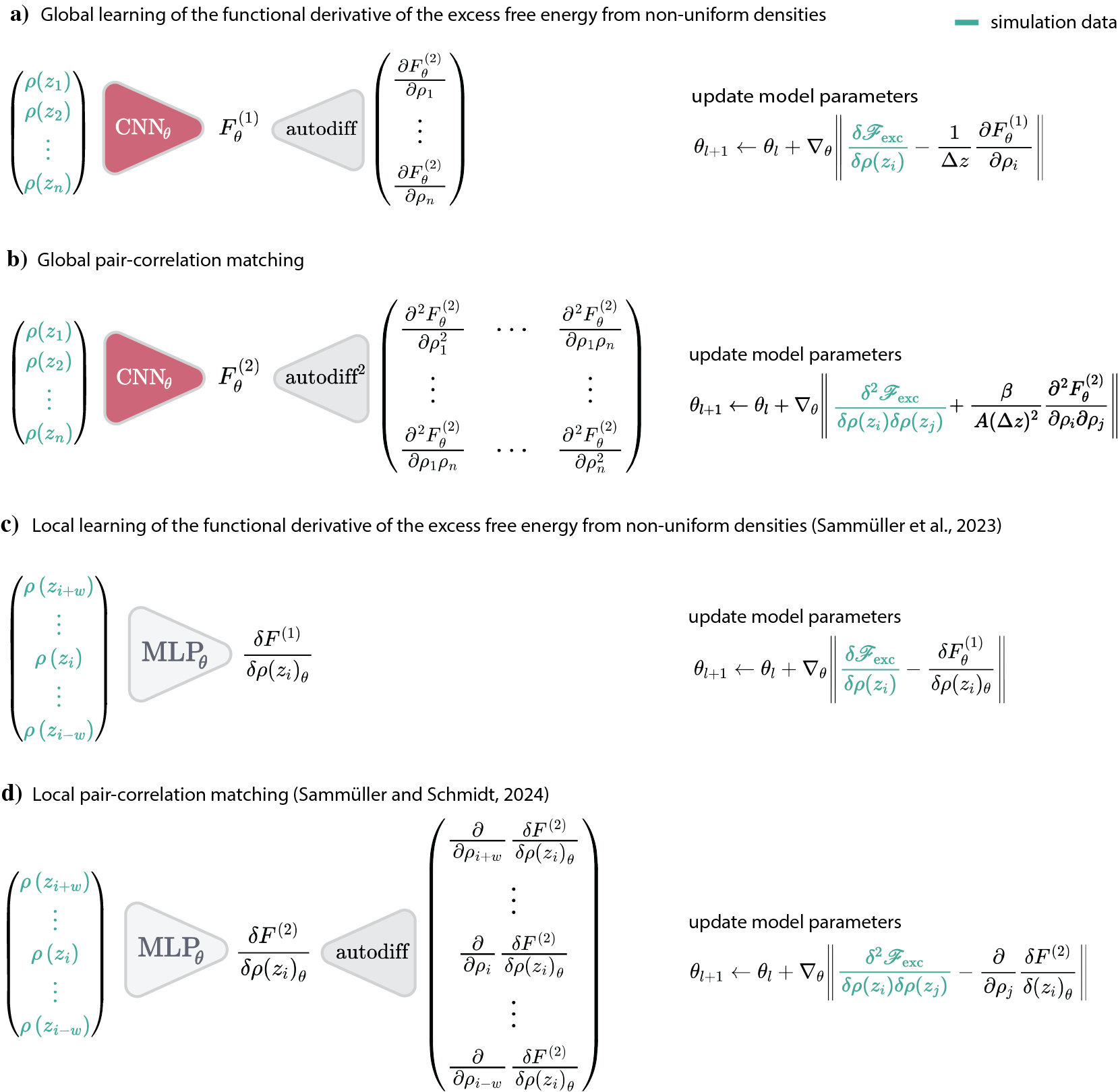}
\caption{\small Overview of the global neural free-energy functionals introduced in this work and the local neural functionals introduced by \textcite{sammullerNeuralFunctionalTheory2023a, sammullerNeuralDensityFunctionals2024}. \textbf{a)} a neural free energy functional trained by fitting the gradient of a convolutional neural network to the first functional derivative of the excess free energy obtained from Monte Carlo simulations, as introduced in this work. \textbf{b)} a neural free energy functional trained with \textit{global} pair-correlation matching, as introduced in this work. \textbf{c)} a local neural functional $\delta F^{(1)}/\delta \rho(z_i)_\theta$ trained by fitting the output of the neural network to the first functional derivative of the excess free energy, as obtained from Monte Carlo simulations. As introduced by \textcite{sammullerNeuralFunctionalTheory2023a}, this approach uses a multi-layer perceptron (MLP) type neural network to estimate $\delta \mathscr{F}_{\text{exc}}/\delta \rho(z_i)$ at position $z_i$ from a local range of density $[\rho(z_{i-w}), \dots, \rho(z_{i}), \dots \rho(z_{i+w})]$ around position $z_i$. \textbf{d)} pair-correlation matching applied in the local learning scheme \cite{sammullerNeuralDensityFunctionals2024}. Here, a neural network approximation for the functional derivative of the excess free energy $\delta F^{(2)}/\delta \rho(z_i)_\theta$ is optimized using a local adaptation of pair-correlation matching. Whereas pair-correlation matching introduced in this work fits the Hessian of the neural free-energy to the second functional derivative of the excess free-energy, here the gradient of the neural functional $\delta F^{(2)}/\delta \rho(z_i)_\theta$ is used to approximate a local range of $\delta^2 \mathscr{F}_{\text{exc}}/\delta\rho(z_i)\delta\rho(z_j)$ for $j = i - w, \dots, i + w$. }
\label{fig:local_learning}
\end{figure*}

\begin{figure*}
    \begin{center}
    \includegraphics[width=1\textwidth]{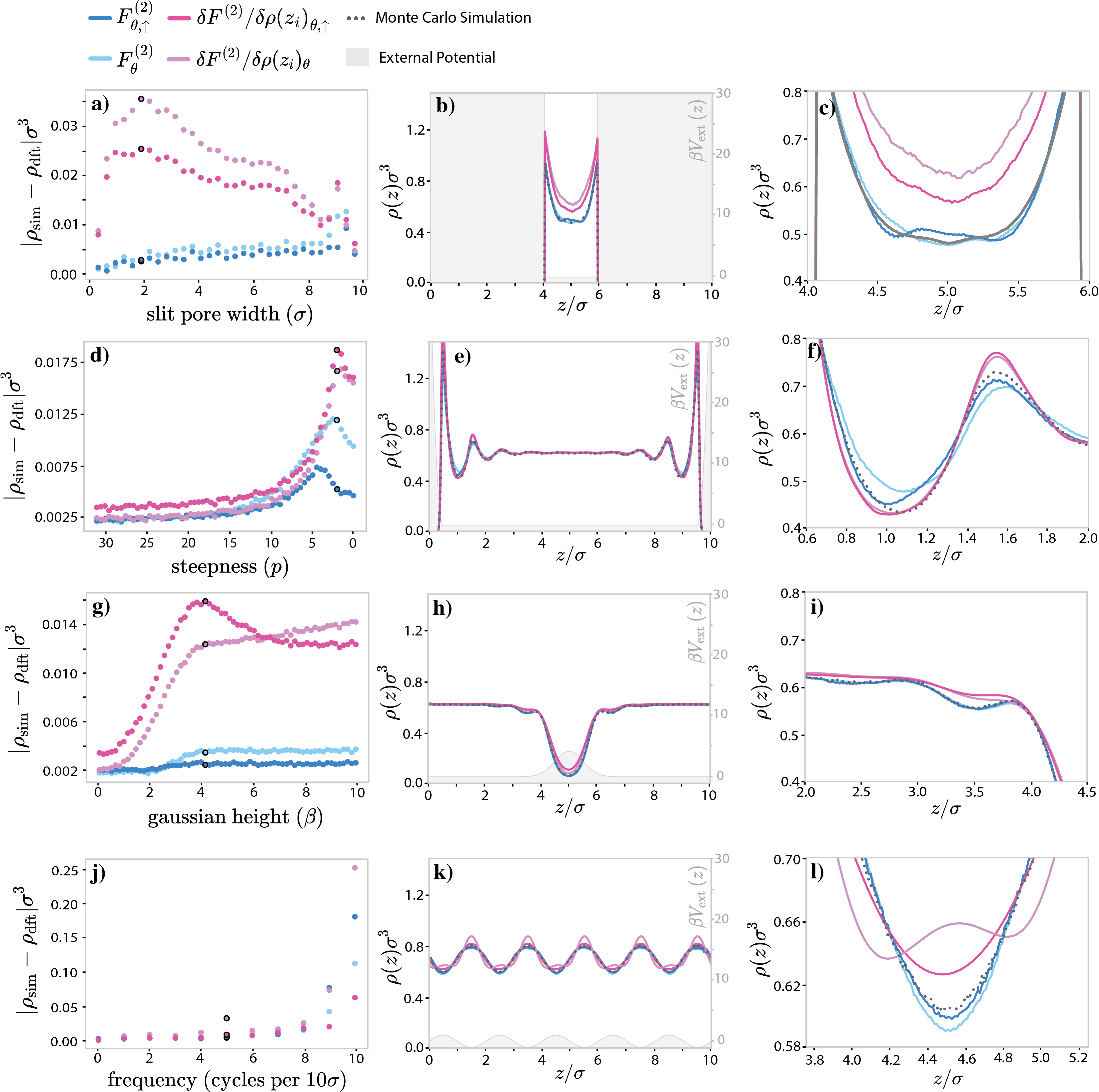}
    \end{center}
    \caption{\small A comparison of the performance of neural functionals $\delta F^{(2)}/\delta \rho(z_i)_\theta$ trained with a local adaptation of pair-correlation matching against neural free-energy functionals $F^{(2)}_\theta$ trained with the global pair-correlation approach on a selection of external potentials with a large degree of inhomogeneity, specifically selected to highlight settings where neural functional predictions deviate from Monte Carlo data when increasing the inhomogeneity. All systems are subjected to $\beta \mu = 0$. Neural functional $\delta F^{(2)}/\delta \rho(z_i)_\theta$ (light purple) is trained with bulk pair-correlation functions in the range $-4 < \beta \mu < 0.5$; Neural functional $\delta F^{(2)}/\delta \rho(z_i)_{\theta,\uparrow}$ (dark purple) is trained with bulk pair-correlation functions in the range $-4 < \beta \mu < 3$; Neural functional $F_{\theta}^{(2)}$ (light blue) is trained with bulk pair-correlation functions in the range $-4 < \beta \mu < 0.5$; Neural functional $F_{\theta, \uparrow}^{(2)}$ (dark blue) is trained with bulk pair-correlation functions in the range $-4 < \beta \mu < 3$. \textbf{a/d/g/j)} mean absolute error $\frac{1}{n} \sum\left|\rho_{\text {sim}}\left(z_i\right)-\rho_{\text{dft}}\left(z_i\right)\right|$ of DFT predictions and MC simulations of particle densities for respectively a slit-pore potential with increasing width; a wall potential with increasing steepness indicated by $p$ as specified in \ref{eq:well}; a Gaussian potential with increasing height; a sine potential with increasing frequency. Errors corresponding to the examples shown in the two rightmost columns are highlighted with a black circle. \textbf{b/e/h/k)} comparison of DFT predictions at samples with a large error of the $\delta F^{(2)}/\delta \rho(z_i)_{\theta,\uparrow}$ functional at respectively a width of $1.9\sigma$; steepness parameter $p=2.1$; Gaussian height of $4.2/\beta$; sine frequency of 0.5 cycles per $1\sigma$. \textbf{c/f/i/l)} Close-up of the most erroneous regions of the DFT estimate of the $\delta F^{(2)}/\delta \rho(z_i)_{\theta,\uparrow}$ functional the system shown in b/e/h/k).}
    \label{fig:local_vs_global_pcm}
\end{figure*}

In \textcite{sammullerNeuralFunctionalTheory2023a}, a local learning approach was proposed for learning the one-body direct correlation function directly. Since $c^{(1)}(x,y,z) = - \beta \delta \mathscr{F}_{\text{exc}}/\delta \rho(x,y,z)$, this is equivalent to learning the first functional derivative of the excess free energy directly. In this approach, the neural functional maps a local neighbourhood of density $[\rho(z_{i-w}), \dots, \rho(z_{i}), \dots \rho(z_{i+w})]$ around position $z_i$ to a single value $\delta \mathscr{F}_{\text{exc}}/\delta \rho(z_i)$ at position $z_i$, as illustrated in Fig. \ref{fig:local_learning}c. This \textit{local} neural functional is trained on $\delta \mathscr{F}_{\text{exc}}/\delta \rho(z_i)$ obtained from sampled inhomogeneous densities, according to Eq.~\ref{eq:dFdrho}. We will indicate this neural functional by $\delta F^{(1)}/\delta \rho(z_i)_\theta$. These notations and illustrations are adapted from \textcite{sammullerNeuralFunctionalTheory2023a} for consistency with the rest of this work and we refer to the original work for further details on this approach. 

This approach is different from the $F^{(1)}_\theta$ functional introduced in this work in two respects. Firstly, $F^{(1)}_\theta$ is a neural functional for the excess free energy whereas the method of \textcite{sammullerNeuralFunctionalTheory2023a} learns a neural functional for the \textit{functional derivative} of the excess free energy, as illustrated in Fig.~\ref{fig:local_learning}. Therefore, the $F^{(1)}_\theta$ functional is optimized by fitting the \emph{gradient} of the neural network output $(1/\Delta z) \partial F_{\theta}^{(1)}/\partial \rho_i$ to $\delta \mathscr{F}_{\text{exc}}/\delta \rho(z_i)$, whereas $\delta F^{(1)}/\delta \rho(z_i)_\theta$ is optimized by fitting the neural network output to $\delta\mathscr{F}_{\text{exc}}/\delta \rho(z_i)$. Secondly, $F^{(1)}_\theta$ is a \textit{global} functional, meaning that it accepts the full density field $[\rho(z_1), \dots, \rho(z_n)]$ as input and estimates $[\mathscr{F}_{\text{exc}}/\delta \rho(z_1),\dots,\mathscr{F}_{\text{exc}}/\delta \rho(z_n)]$ for the full system. In contrast, $\delta F^{(1)}/\delta \rho(z_i)_\theta$ is a \textit{local} functional, which takes as input a local neighborhood of density $[\rho(z_{i-w}), \dots, \rho(z_{i}), \dots \rho(z_{i+w})]$ and estimates $\delta\mathscr{F}_{\text{exc}}/\delta \rho(z_i)$ at position $z_i$.

Similar to the local approach for learning $\delta\mathscr{F}_{\text{exc}}/\delta \rho(z_i)$ from inhomogeneous densities, pair-correlation matching can also be performed within this local learning scheme, as introduced by \textcite{sammullerNeuralDensityFunctionals2024} in response to an earlier version of this work. Here, the objective is to approximate $\delta^2 \mathscr{F}_{\text{exc}}/\delta \rho(z_i) \delta \rho(z_j)$ for $j \in 1, \dots, n$ by the gradient of the neural functional $\delta F^{(2)}/\delta \rho(z_i)_\theta$, as illustrated in Fig. \ref{fig:local_learning}d. This is different from the \textit{global} pair-correlation matching approach used in this work, where the neural functional $F^{(2)}_\theta$ takes as input the density in the entire system $[\rho(z_1), \dots, \rho(z_n)]$, and is optimized by fitting the Hessian of the neural network to $\delta^2 \mathscr{F}_{\text{exc}}/\delta \rho(z_i) \delta \rho(z_j)$ for all $i, j \in 1, \dots, n$, with $n$ as the total number of gridpoints across the system. 

As is evident in the results by \textcite{sammullerNeuralDensityFunctionals2024}, the local version of pair-correlation matching does not seem to match the predictive capabilities of the global pair-correlation method introduced in this work. Here we implement this local version of pair-correlation matching to further investigate this discrepancy. 

We implement similar \textit{local} neural functionals to the approach by \textcite{sammullerNeuralFunctionalTheory2023a}, a multi-layer perceptron (MLP) with 3 hidden-layers, each with 512 nodes. The neural network takes as input bulk densities of resolution $\Delta z = \sigma/100$. The input of the neural network is a local window of bulk densities $[\rho_b(z_{i-w}), \dots, \rho_b(z_i), \dots, \rho_b(z_{i+w})]$, a window of $3.5\sigma$ on both sides of $z_i$, meaning $w=350$ gridpoints on both sides of $\rho(z_i)$. We train $\delta F^{(2)}/\delta \rho(z_i)_{\theta}$ on bulk pair-correlation functions in range $-4 < \beta \mu < 0.5$, and $\delta F^{(2)}/\delta \rho(z_i)_{\theta, \uparrow}$ on pair-correlation functions in range $-4 < \beta \mu < 3$. We compare with \textit{global} neural functional $F^{(2)}_\theta$ trained on pair-correlation functions in range $-4 < \beta \mu < 0.5$ and $F^{(2)}_{\theta,\uparrow}$, trained in range $-4 < \beta \mu < 3$. We expand upon training neural functionals with pair-correlation matching in range $-4 < \beta \mu < 3$ in Section 9. Both $F_\theta^{(1)}$ and $F_\theta^{(2)}$ are trained on data with grid-spacing $\Delta z = \sigma/100$ as well, as detailed in Section 10. 

We compare the accuracy of local and global neural functionals trained using pair-correlation matching on the same set of highly inhomogeneous external potentials as discussed in Section 7. We see that across all external potentials investigated, the local version of pair-correlation matching yields higher errors and a larger error increase with increase in inhomogeneity of the potential, as shown in Fig. \ref{fig:local_vs_global_pcm}. At this stage it is not clear what is the reason behind this discrepancy in predictive power between pair-correlation matching applied to global and local learning schemes. 


\section{9. Extending the Training Set for Neural Functionals}

Since the pair-correlation matching approach relies on the inhomogeneity of the pair-correlation function for $F_\theta^{(2)}$ to approximate inhomogeneous densities, it is reasonable to question whether including pair-correlation functions at higher bulk densities in the training set would increase the performance of the pair-correlation matching approach for predicting highly inhomogeneous densities. Therefore, the results of Section 7 and Section 8 include a comparison with neural functional $F_{\theta,\uparrow}^{(2)}$, which has been trained on direct correlation functions in range $-4 < \beta \mu < 3$. This is an extension of the train set of the neural functional $F_\theta^{(2)}$ for which the train set sits within the range $-4 < \beta \mu < 0.5$, as detailed in the main paper. The train set of $F_{\theta,\uparrow}^{(2)}$ consisted of 1000 direct correlation functions obtained from radial distribution functions sampled in cubic systems of size $(10\sigma)^3$, as well as 134 direct correlation functions in the range $0.5 < \beta \mu < 3$ obtained from from cubic systems of size $(20\sigma)^3$. 

To obtain $c^{(2)}_b(r)$ through Eq.~8 of the main paper, we assume that $g(r)$ has converged to unity at distance of half the box size, $r=L/2$. For the cubic systems with edge length $L=10\sigma$, we found this condition to hold for bulk densities $\rho_b\sigma^3<0.67$ corresponding to $\beta \mu < 0.5$; for cubic systems with edge length $L=20\sigma$, we found the condition to hold for bulk densities $\rho_b\sigma^3<0.82$ corresponding to $\beta \mu < 3$. 

Since the direct correlation function is short ranged \cite{hansenTheorySimpleLiquids2013}, we combine the direct correlation functions obtained from systems of size $(20\sigma)^3$ by cutting the tail of the computed $c^{(2)}(r)$ at $5\sigma$, where $c^{(2)}(r) \rightarrow 0$, such that all computed $c^{(2)}(r)$ have range $0 \leq r \leq 5\sigma$ and are all used to compute $\delta^2 \mathscr{F}_{\text{exc}}/\delta \rho(z_i) \delta \rho(z_j)|_{\rho_b}$ within a planar system of size $L=10\sigma$. In bulk, the range $-4 < \beta \mu < 3$ corresponds to $0.02 < \rho_b\sigma^3 < 0.82$, exceeding the maximum bulk density $\rho_b\sigma^3 = 0.67$ for the range $-4 < \beta \mu < 0.5$.

In Section 8, we additionally compare with neural functional $F_{\theta,\uparrow}^{(1)}$, for which the train set of inhomogeneous densities similarly been extended from samples in the range $-4 < \beta \mu < 0.5$ to $-4 < \beta \mu < 3$. The dataset consists of 800 train samples. The inhomogeneous densities contained in this extended dataset are induced by the same type of external potentials as in the rest of this work, as detailed in Section 4.

These neural functionals $F_{\theta,\uparrow}^{(1)}$ and $F_{\theta,\uparrow}^{(2)}$ were both trained at a resolution of $\Delta z = \sigma/100$ and the neural network architecture detailed in Section 11.


\section{10. Neural Functionals with Increased Resolution}

The detailed comparisons of Sections 7 and 8 were performed with neural functionals $F_\theta^{(1)}$ and $F_\theta^{(2)}$ that were trained with respectively inhomogeneous densities and bulk direct correlation functions of grid-spacing $\Delta z = \sigma/100$, and therefore estimate densities with this resolution of $\Delta z = \sigma/100$ as well. This is an increase in resolution of the training data from a grid-spacing of $\sigma/32$, as used in the main paper. Together with this increase in resolution, the number of layers in the convolutional neural network of the neural functionals $F_\theta^{(1)}$ and $F_\theta^{(2)}$ was increased from 6 to 8, with the number of channels per layer as $N_{\text{channels}} =[32,32,32,32,64,64,64,64]$, applying average-pooling with kernel size 2 after each layer. These adaptations were applied to optimize the predictive performance of these functionals and as an attempt to limit the effect of numerical errors, such as accumulated in the numerical transformation from the $g(r)$ to $\bar{c}^{(2)}(|z-z'|)$, on the estimates of the neural functionals. Additionally, we changed the parameter $\alpha$ in the Picard iteration \cite{rothIntroductionDensityFunctional2006, edelmannNumericalEfficientWay2016, mairhoferNumericalAspectsClassical2017} from 0.1 to 0.01 for better convergence. Lastly, the loss factors $\alpha=1/1000$ and $\beta=1/32$ of the $F_\theta^{(2)}$ functional were changed to $\alpha=1$ and $\beta=1$ (see the End Matter of the main paper), to place more importance on the correct offset of the predicted density. This slightly reduced the error of the $F^{(2)}$ functional for lower densities. 

All neural functionals $F_{\theta}^{(1)}$, $F_{\theta}^{(1)}$, $F_{\theta,\uparrow}^{(1)}$ and $F_{\theta,\uparrow}^{(2)}$ as discussed in Sections 7 and 8, were trained with these adaptations for higher resolution neural functionals. The neural functional $F_{\theta}^{(1)}$ with increased resolution was trained on 800 inhomogeneous densities in range range $-4 < \beta \mu < 0.5$. The $F_{\theta}^{(2)}$ functional with increased resolution was trained on 1000 bulk direct correlation functions in range $-4 < \beta \mu < 0.5$. This difference in dataset size was due to the fact that no bulk direct correlation functions had to be held in the test set, as the performance of the neural functional was validated on inhomogeneous densities instead. 

The neural functional $F_{\theta, \uparrow}^{(1)}$ with increased resolution was trained on 800 inhomogeneous densities in range range $-4 < \beta \mu < 3$. The $F_{\theta, \uparrow}^{(2)}$ functional with increased resolution was trained on 1000 direct correlation functions in range $-4 < \beta \mu < 0.5$ and 134 direct correlation functions in the range $0.5 < \beta \mu < 3$. These functionals are further detailed in Section 9.


\section{11. Sampling Inhomogeneous Densities in Planar Geometry Systems vs. Arbitrary 3D Systems}

\begin{figure}
    \begin{center}
    \includegraphics[width=0.35\textwidth]{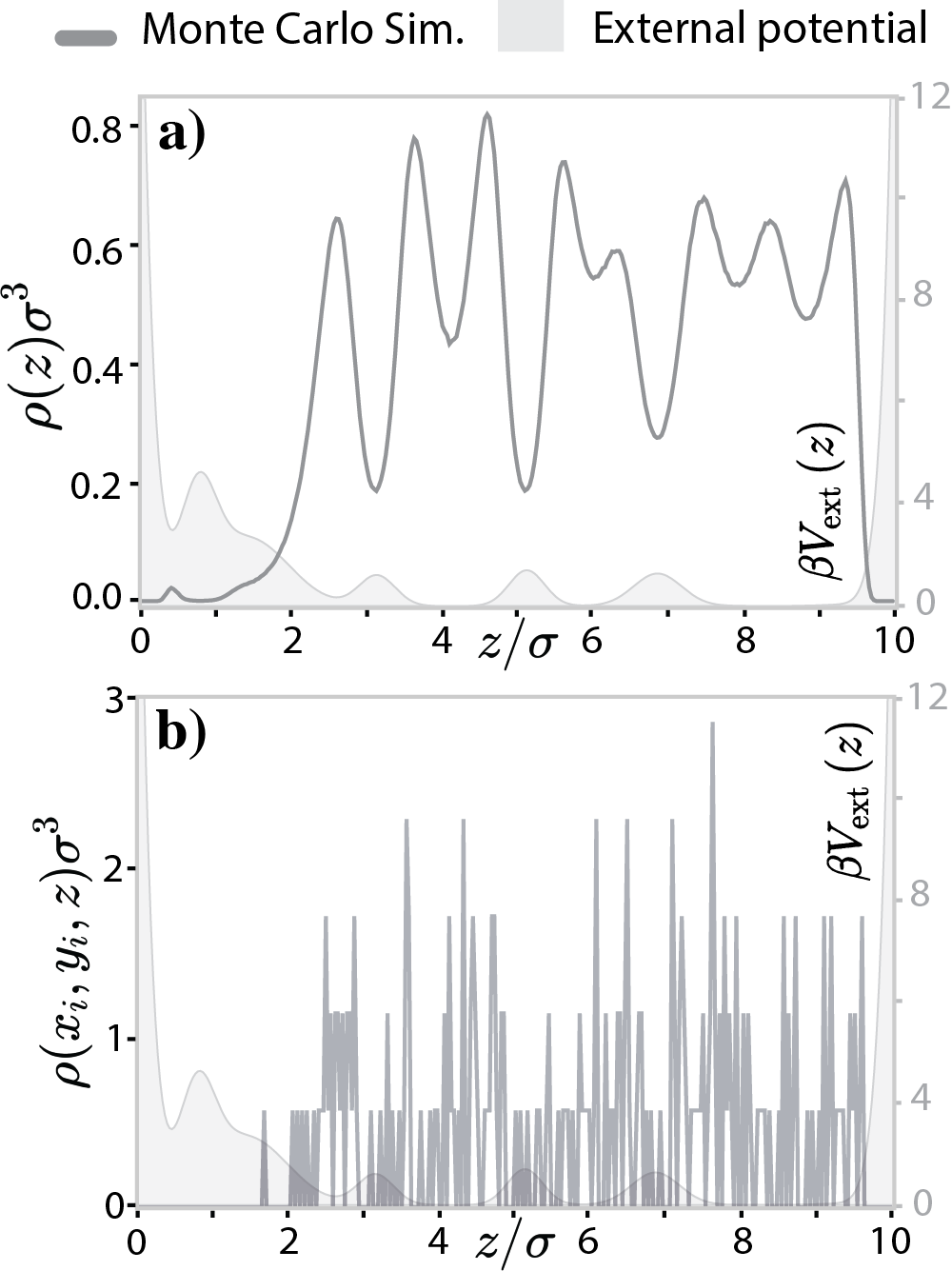}
    \end{center}
    \caption{\small Comparison between sampling densities $\rho(z)$ in planar geometry and $\rho(x,y,z)$ in arbitrary three-dimensional geometry. \textbf{a)} $\rho(z)$ in an external potential $V_{\text{ext}}(z)$ (shown in gray) sampled from a MC simulation with $10^9$ trial moves. \textbf{b)} Slice of $\rho(x,y,z)$ at $x_i$ for the same laterally symmetric external potential $V_{\text{ext}}(z)$, also sampled from a MC simulation with $10^9$ trial moves.}
    \label{fig:sampling3D}
\end{figure}

The density profiles in planar geometry used in this study require approximately $1$ hour of CPU time per density profile. Extrapolating these times naively to full three-dimensional density profiles, it would require an impractical $A/(\Delta x \Delta y) \cdot 1 \text{ hours} = 102400$ hours to generate each density with similar accuracy for a resolution of $\Delta x = \Delta y = \Delta z = \sigma/32$ as used in this work. To illustrate this, we compare a 1D slice from a 3D density profile with a 1D density profile in planar geometry, both sampled within the same number of MC steps ($10^9$ trial moves with $10^7$ equilibration moves and 4 decorrelation cycles) (Fig.~\ref{fig:sampling3D}). These density profiles are constructed by binning particle positions in intervals of $4\cdot N_{particles}$ trial moves throughout the MC simulation. After completion of the simulation, only a very low number of particles has been counted in each bin of the 3D histogram that constructs the density of Fig.~\ref{fig:sampling3D}b. Many bins remain empty and many contain only  a few particles, creating the discrete peaks of Fig.~\ref{fig:sampling3D}b. 

This illustrates that much longer sampling times are necessary to sample accurate three-dimensional density profiles. The results presented in this paper provide a compelling alternative: the radial distribution functions  sampled for this work were already obtained from 3D bulk systems, after which they were numerically transformed into direct correlation functions in planar geometry. This means that it is likely that the same dataset of radial distribution functions can be used when extending this approach to arbitrary three-dimensional systems, with the only difference that a numerical transformation to the \textit{radially symmetric} pair-correlation function $c^{(2)}(r)$ needs to be applied.

\bibliography{references}